\documentclass[10pt]{iopart}
\usepackage{iopams}  
\usepackage{graphicx}
\usepackage{epstopdf}
\usepackage{cite}
\begin{document}

\title[Graphene on metals]{Graphene growth and properties on metal substrates}

\author{Yuriy Dedkov}

\address{SPECS Surface Nano Analysis GmbH, Voltastrasse 5, 13355 Berlin, Germany}
\ead{Yuriy.Dedkov@specs.com}

\author{Elena Voloshina}

\address{Institut f\"ur Chemie, Humboldt-Universit\"at zu Berlin, 10099 Berlin, Germany}
\ead{Elena.Voloshina@hu-berlin.de}
\vspace{10pt}

\begin{abstract}
Graphene-metal interface as one of the interesting graphene-based objects attracts much attention from both application and fundamental science points of view. This paper gives a timing review of the recent experimental works on the growth and the electronic properties of the graphene-metal interfaces. This work makes a link between huge amount of experimental and theoretical data allowing to understand the influence of the metallic substrate on the electronic properties of a graphene overlayer and how its properties can be modified in a controllable way. The further directions of studies and applications of the graphene-metal interfaces are discussed. 
\end{abstract}

\pacs{73.20.-r, 73.22.Pr, 74.55.+v, 79.60.-i}

\vspace{2pc}
\noindent{\it Keywords}: graphene, metallic surfaces, DFT, NEXAFS, ARPES, LEEM, PEEM, STM, AFM

\submitto{\JPCM}

\maketitle

\ioptwocol

\section{Introduction}

\begin{figure}[t]
\includegraphics[width=\linewidth]{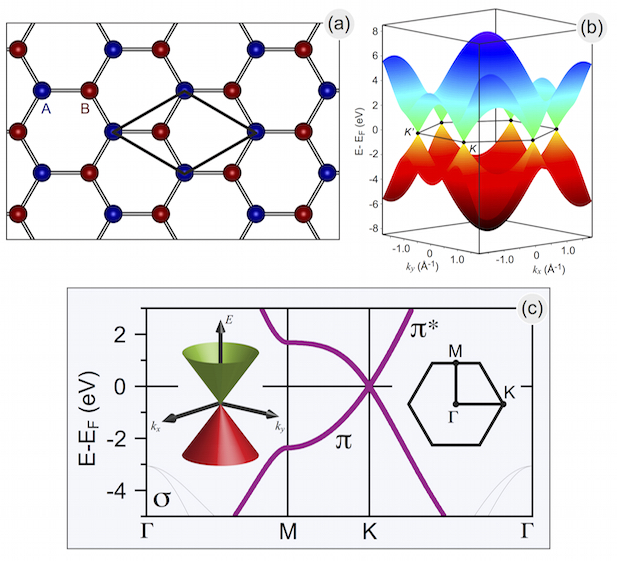}
\caption{(a) Crystallographic structure of graphene with two equivalent carbon atoms, A and B, in the unit cell marked by the solid-line black rhombus. (b) Two-dimensional electronic structure of graphene [$E(k_x,k_y)$] in the tight-binding approximation~\cite{CastroNeto:2009}. (c) The DFT calculated band structure of graphene in the vicinity of $E_F$ along the main directions in the hexagonal Brilloine zone (shown as an inset)~\cite{Voloshina:2012c}. The dispersion of $\pi$ and $\pi^*$ states is highlighted by the thick line.}
\label{gr_str-and-bands}
\end{figure}

Graphene (gr), a two-dimensional layer of carbon atoms arranged in a honeycomb lattice [Fig.~\ref{gr_str-and-bands}(a)], demonstrates a variety of unique electronic and transport properties. The discovery of such fascinating phenomena as very high electron and hole mobilities, ambipolar electric field effect, integer and half-interger quantum Hall effects for electrons and holes, etc.~\cite{Novoselov:2005,Zhang:2005} attracts a lot of attention from the different fields of solid state physics and chemistry.

The hexagonal lattice of graphene consists of two equivalent carbon sublattices [A and B in Fig.~\ref{gr_str-and-bands}(a)]. Carbon atoms in this structure are $sp^2$ hybridzed (one $2s$ and two $2p$ orbitals), that leads to the formation of the planar structure with strong $\sigma$ bonds and the distance of $1.42$\,\AA\ between carbon atoms. Completely filled $\sigma$ bands for the infinite graphene layer forms deep valence band levels of graphene. The $2p_z$ orbitals of the neighbouring carbon atoms are perpendicular to the plane of graphene and their overlap above and below this plane leads to the formation of the $\pi$ valence band. This band is half filled and the unique property of the electronic structure of graphene is that the $\pi$ and $\pi^*$ bands touch at the corners of the hexagonal Brillouine zone of graphene (at the $K$ points) directly at the Fermi level ($E_F$) [Fig.~\ref{gr_str-and-bands}(b,c)]. The band dispersion, $E(k)$, of the $\pi$ states in the close vicinity of $E_F$ is linear and described via the Dirac equation $E=\hbar v_F k$, where $v_F\approx 1\times10^6$\,m/s is the Fermi velocity (``speed of light'' for the massless Dirac fermions). The point where $\pi$ bands intersect at $K$ is called a Dirac point. Thus, graphene is a semimetal in the undoped case. Due to the low density of the valence band states (DOS) of graphene around $E_F$ and the linear dependence of DOS on energy, the electronic structure of graphene can be strongly modified in a controllable way via attaching different species to graphene that can lead to different technological applications of graphene, like touch screens or gas sensors~\cite{Schedin:2007,Bae:2010,Ryu:2014fo}.

First transport experiments on graphene were performed on the single graphene flakes exfoliated through the so-called ``scotch-tape'' method from bulk graphite. However, despite the high crystalline order and transport mobility of these flakes, they cannot be used in the technological process due to their small size and uncontrollable preparation procedure. In a series of experiments it was shown that graphene layers of very high quality can be prepared on SiC semiconducting substrates~\cite{Norimatsu:2014be}. These graphene layers on SiC demonstrate transport properties comparable to those for the exfoliated graphene~\cite{Emtsev:2009,Miller:2009} and it was shown that the electronic properties of graphene/SiC interface can be tailored in a wide range of properties via adsorption or intercalation of different species~\cite{Norimatsu:2014be}. Later it was demonstrated that graphene nanoribbons can be prepared on the templated SiC substrate using scalable photolithography and microelectronics processing and this technology leads to the fabrication of the graphene transistors with density of 40,000/cm$^2$~\cite{Sprinkle:2010}. 

However, as was shown in the recent works the most promising method for the preparation of the graphene layers, which can be further used in the technological applications, is the synthesis of single- and multilayer graphene on the metallic support via chemical vapour deposition (CVD). This method combined with roll-to-roll method allows production of the graphene layers on metallic substrate and then transfer on the polymer support up to meter sizes~\cite{Bae:2010}. In the end of 2009 the full technological process was presented that leads to the fabrication of the first graphene-based touch screens. Later, the optimisation of this technology allowed to fabricate the first mobile phone with the screen fully made on the basis of graphene~\cite{Ryu:2014fo}. These graphene layers demonstrate the electron mobility of $5100-5200$\,cm$^2$V$^{-1}$s$^{-1}$ at room temperature, visible light transparency of 97.75\% at 550\,nm, and quantum Hall effect~\cite{Bae:2010,Ryu:2014fo}.

Graphene (known as monolayer of graphite in a former time) overlayers on metallic surface are in focus of surface science research for many years starting from middle of the 60s~\cite{Hagstrom:1965vh,Lyon:1967,May:1969uj,Grant:1970,Wintterlin:2009,Batzill:2012,Dedkov:2012book}, when the multidomain graphene layer on Pt(100)~\cite{Hagstrom:1965vh} and the moir\'e graphene structure on Ru(0001)~\cite{Grant:1970} were identified in the low-energy electron diffraction (LEED) experiments. At that time the main interest to this topic was motivated by the studies of the catalytic activity of the clean $d$-metal surfaces and in most cases graphene was considered as a poison layer blocking their reactivity.

The discovery of the fascinating transport properties of graphene renewed interest to the investigation of the graphene/metal interfaces. Here the studies are ranged from the fundamental problems, like the correct description of the relatively weak interaction between graphene and metal surface, which in most cases leads to the drastic modification of the electronic structure of graphene, to the more practical issues, like the preparation of the ordered arrays of clusters on the graphene moir\'e on the surface of $4d$ and $5d$ metals, which then can be used as a model system for the studies of e.\,g. catalytic properties of these systems.  

The present manuscript gives a timing review of the recent experimental and theoretical results on the graphene/metal interfaces. Starting from the consideration of the methods of graphene's growth on metallic surfaces, it follows with the overview of the crystallographic and electronic properties of these systems, including the experiments on the modification of graphene/metal via adsorption and intercalation of different species and formation of graphene nano-objects. This review includes several examples of the representative results obtained with the main surface science methods (STM, AFM, LEEM-PEEM, NEXAFS, XPS, ARPES) and tries to make links between them with the help of the state-of-the-art density functional theory (DFT) calculations.

\begin{figure*}[t]
\includegraphics[width=\linewidth]{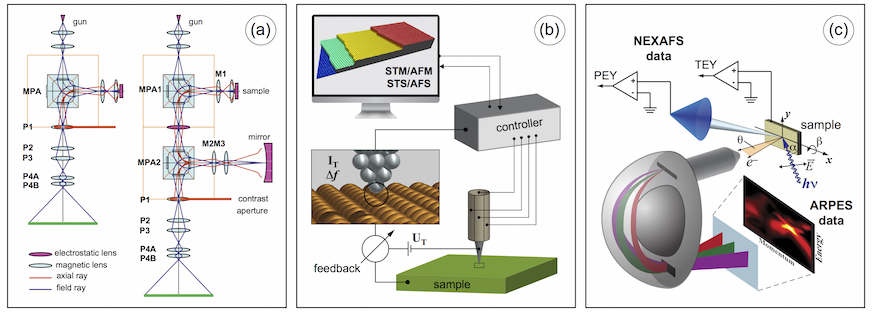}
\caption{Principal schemes of (a) low-energy electron microscopy (figure is taken from Ref.~\cite{Tromp:2010hj} with permission), (b) scanning probe microscopy/spectroscopy (STM/AFM), and (c) photoelectron spectroscopy (XPS/ARPES) and near-edge x-ray absorption spectroscopy (NEXAFS/XMCD) experiments.}
\label{exp-methods}
\end{figure*}

\section{Experimental methods}

Graphene layers on metallic surfaces are pure 2D objects and they are ideal objects for the application of different surface science techniques. Low-energy electron microscopy (LEEM) and scanning probe microscopy (SPM) methods give information about morphology of the system and the electronic structure on the local scale (from $\mu$m to atomic scale). The photoelectron spectroscopy based techniques (NEXAFS and PES) allow to get information about the electronic structure on the macroscopic scale (from several hundreds nm to mm scale). This section gives a short overview of the listed techniques pointing their (dis-)advantages. 

\subsection{LEEM/PEEM}

LEEM is a parallel imaging technique, which uses elastically backscattered electrons to image crystalline surfaces and interfaces [Fig.~\ref{exp-methods}(a)]. Due to the large electron backscattering cross section of most materials the intensities of the diffracted and/or reflected beams are quite high, that makes LEEM the ideal technique to image in video rate dynamic processes such as surface reconstructions, epitaxial growth, step dynamics, self-organization and others. LEEM enables to reach a lateral resolution of less than 10 nm. Most importantly, it offers several structure sensitive complementary imaging and diffraction methods to probe the surface. The detailed description of the operation principles of this method can be found elsewhere~\cite{Bauer:1994wt,Tromp:2000wt,Tromp:1998ua,Tromp:2010hj}.

A beam of high energy electrons ($10-20$\,keV), decelerated by the retarding potential of the objective lens, impinges in normal incidence on the surface, with energy in the range 0 to few hundreds eV. The beam energy is varied by adjusting a bias voltage between sample and the emitter. The elastically backscattered electrons are then reaccelerated through the objective lens, following the inverse pathway. The objective produces a magnified image of the surface in the beam separator, which is further magnified by several additional lenses in the imaging column of the instrument. This image is projected onto an imaging detector with microchannel plate and phosphorous screen, and finally acquired by a CCD camera.  Along with the real space imaging, a LEEM microscope is also capable of the reciprocal space imaging.

Depending on the diffracted beam selection either bright field of dark field imaging in LEEM is possible. If the zero-order diffracted beam (``00'' beam) is selected by a contrast aperture positioned in the diffraction plane then the bright-field LEEM is performed (defocusing of the beam allows to convert the phase difference, appeared due to the reflection from different heights, into an amplitude difference that gives a possibility to image steps at the surface). In this case the pure structural contrast is obtained. In case of overlayers on the surface, the interference of the backscattered beams from the surface and interface can produce the maxima and minima in the intensity that allows to measure a local thickness of the overlayer. If the secondary diffracted beam is selected then a dark-field image of the surface is produced. Here all areas that contribute to the formation of the selected beam appear bright. 

In mirror electron microscopy (MEM) the surface is illuminated with electrons at very low energy, so that the electrons interact very weakly with the surface (this occurs at the transition MEM-LEEM). Under these conditions the contrast is due to work function differences and topography variations. MEM allows non-crystalline samples to be imaged.

In case if photon source is used instead of the electron gun, a LEEM can be used for the imaging of photoemitted electrons. If the entrance slit is placed on the entrance of the the prism array (MPA) on the way of the emitted electron than the use of the reciprocal imaging produce the $k$-space resolved photoelectron intensity map allowing to perform local imaging of the electronic structure.   

\subsection{Scanning probe microscopy methods}

SPM methods combined with the corresponding spectroscopy add-ons give an information about morphology of the studied objects and the electronic structure down to the atomic scale [Fig.~\ref{exp-methods}(b)]~\cite{Hofer:2003wk,Giessibl:2003aaa,Chiang:2011aaa}. In scanning tunnelling microscopy (STM) the feedback is regulated via setting the value of the tip position above the surface, $z$, that regulates the tunnelling current, $I_T$, which exponentially depends on $z$, that gives the distribution of the local density of states at the surface, $\rho(x,y,z)$. Here $I_T(x,y,z,V)\propto \int \rho(x,y,z,E)dE$, where the integration is performed in the energy range between the Fermi levels ($E_F$) of tip and sample, when they differ by the value $eU_T$, where $U_T$ is the bias voltage. One can perform the differentiation of the $I(x,y,z,U_T)$ signal with respect to the bias voltage (scanning tunnelling spectroscopy, STS). This can be carried out with lock-in technique and gives a direct information about the local density of states. If such measurements performed in the scanning mode at different bias voltages, then it allows to observe the so-called electron density standing waves of different periodicities that can be used to obtain the electron dispersion relation, $E(k)$, of the surface electronic states~\cite{Davis:1991wn,Hormandinger:1994vt,Burgi:2000ux}.

In the modern atomic force microscopy (AFM) experiments, the oscillating conducting tip is used. Approach of the tip to the sample slightly changes the resonance frequency of the sensor, $f_0$, and this frequency shift value, $\Delta f$, can be used as a signal for the feedback loop to map the sample topography. The total interaction energy between tip and surface is a sum of the long-range electrostatic and van der Waals contributions and the short-range chemical interaction which provides the atomic contrast during scanning. As an \textit{add on} to AFM the method of the Kelvin-probe force microscopy (KPFM) was developed~\cite{Melitz:2011}, that allows to obtain information about the local distribution of the contact potential difference between tip and surface ($U_{LCPD}$). In this method the DC and AC (with the frequency $\omega$) voltages ($U_{DC}$ and $U_{AC}$, respectively) are applied between tip and sample and the measured first harmonic of the $\Delta f$ signal is used to nullify the difference between $U_{DC}$ and $U_{LCPD}$.    

\begin{figure*}
\includegraphics[width=\linewidth]{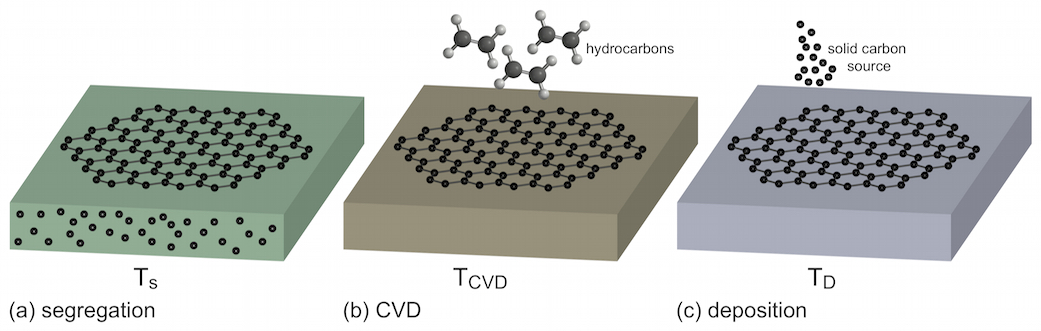}
\caption{Methods of the graphene synthesis via (a) carbon segregation from the metal bulk, (c) CVD from hydrocarbons, and (c) carbon deposition from solid-carbon evaporation sources.}
\label{gr_synthesis}
\end{figure*}

STM and AFM can be used for collecting the 3D $I_T(x,y,z)$ and $\Delta f(x,y,z)$ sets of data. In the first approach the single $I_T(z)$ and $\Delta f(z)$ curves are collected on the dense grid and then they are combined in the 3D sets. In the second case the constant height 2D maps $I_T(x,y)$ and $\Delta f(x,y)$ maps acquired with the usual scanning resolution at different equidistant distances and then combined in a 3D set. Such experiments are time consuming; therefore the careful postprocessing of the data, taking into account the thermal drift of the oscillating sensor and the scanner, is necessary. The detailed discussion of both approaches can be found in the literature~\cite{Albers:2009ig,Baykara:2010a,Dedkov:2014di}.     

\subsection{NEXAFS and XMCD}

The method of near-edge x-ray absorption spectroscopy (NEXAFS) and its respective extension, x-ray magnetic circular dichroism (XMCD), are used to obtain information about energy distribution of the unoccupied valence band states above $E_F$ and for the determination of the magnetic moment of elements in the system [Fig.~\ref{exp-methods}(c)]~\cite{Wende:2004,Stohr:1999a}. In these methods the photon energy is scanned around the particular x-ray absorption threshold and the corresponding total electron yield (TEY mode of NEXAFS) is measured as a drain current from the sample giving information about the absorption coefficient, which is proportional to the density of unoccupied states. If the photoelectron yield is measured by the electron multiplier then the low energy electrons can be removed from the signal by applying the negative potential of several volts to the electrostatic grid placed in front of the detector that allows to increase drastically the surface sensitivity of the method (partial electron yield or PEY). If linear polarized light is used in such experiments, then varying the angle of the impugning light on the sample one can obtain the information about orientation of the valence band orbitals of the species adsorbed on the surface (the so-called \textit{search-light}-like effect)~\cite{Stohr:1999b}. If magnetic sample is studied and circularly polarised light is used, then the absorption coefficient for the x-ray light depends on the relative orientation of the sample magnetisation and projection of the spin of light on this direction (XMCD experiment). If two NEXAFS spectra for opposite directions of magnetisation are measured, then they can be used for the calculation of the spin ($\mu_S$) and orbital ($\mu_L$) magnetic moments of elements in the system~\cite{Thole:1992,Carra:1993}.

\subsection{Photoelectron spectroscopy}

In the method of photoelectron spectroscopy electrons are excited from the core levels or occupied valence band states and are analysed by their kinetic energy ($E_{kin}$) giving a replica of the electronic structure of solids~\cite{Reinert:2005}. Analysis of the core levels (x-ray photoelectron spectroscopy, XPS) gives an information about the chemical states of elements that might be influenced by the corresponding atomic coordination. In case of the angle-resolved photoelectron spectroscopy (ARPES) [Fig.~\ref{exp-methods}(c)] the electrons emitted from the valence band are additionally analysed by their emission angle ($\theta$) and detected by the 2D detector giving photoemission intensity map for one $k$-direction in the Brillouin zone. In the modern ARPES experiments the sample is usually placed on the goniometer having 5 or 6 degrees of freedom. As shown in Fig.~\ref{exp-methods}(c), additional stepwise rotation of a sample around $x$-axis by angle $\beta$ produces a stack of data $I(E_{kin},\theta,\beta)$, which can be transferred to the respective set $I(E_B,k_x,k_y)$, where $E_B$ and $k_{x,y}$ are the binding energy and the respective component of the $k$-vector of electron in the valence band~\cite{Reinert:2005,Damascelli:2003,Damascelli:2004}.

\section{Graphene growth on metals}

There are three main methods of the graphene synthesis which are widely used now (Fig.~\ref{gr_synthesis}): (a) segregation of the dissolved carbon from the bulk metal, (b) chemical vapour deposition (CVD) from hydrocarbons, and (c) deposition from solid carbon source (or molecular beam epitaxy, MBE). Here these methods are discussed in details.

\begin{figure}
\includegraphics[width=\linewidth]{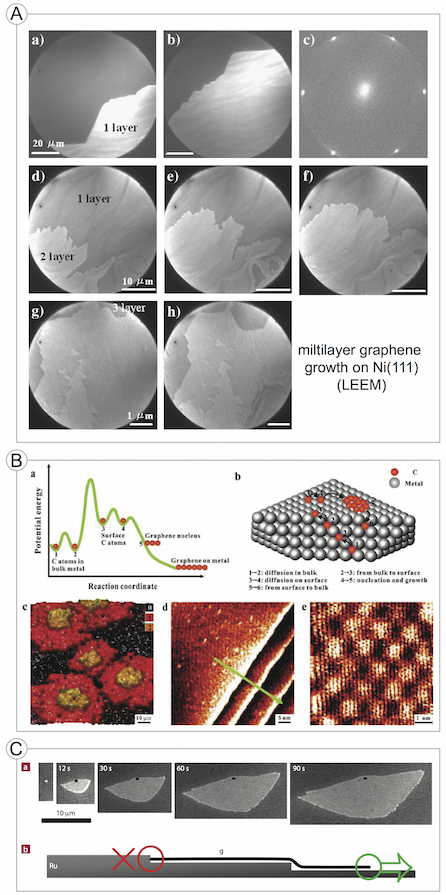}
\caption{(A) LEEM images of the graphene segregation on Ni(111) at different stages: (a,b) the first layer segregates at $1125$\,K, (d-f) the second layer at $1050$\,K and (g-h) the third layer at $1050$\,K. Image (c) is a $\mu$LEED pattern of the single-layer graphene/Ni(111) demonstrating $(1\times1)$ structure. Data are taken from \cite{Odahara:2011iw} with permission. (B) (a,b) Energy profile and the corresponding dynamic processes involved in graphene segregation. (c-e) STM images of graphene at different steps of segregation. Data are taken from \cite{Liu:2011bl} with permission. (C) (a) Real-time LEEM images of graphene segregation on Ru(0001) at $850^\circ$\,C. (b) Schematic illustration of the ``downhill'' graphene growth. Data are reproduced from \cite{Sutter:2008} with permission.}
\label{gr-met_segregation}
\end{figure}

\subsection{Segregation}

This method can be used for the graphene synthesis on metals which have high carbon solubility, like Ni or Ru~\cite{Yu:2008,Liu:2011bl,Liu:2011df,Odahara:2011iw,Dahal:2012jh,Sutter:2008}, but can work also for the metals which are heated to the high temperature that their carbon solubility becomes significant~\cite{Sutter:2009a,Nie:2011fn,Gao:2012bg,Sun:2014gv,Mok:2014kn}. In this case, initially the metal bulk is loaded with carbon atoms at high temperature and then during cooling carbon atoms travel to the surface and form ordered layer of graphene. This method overcomes the main limitations of the CVD method as the multilayer graphene can be easily grown; thickness of a graphene layer depends on the cooling rate as well as on the loading and segregation temperature ($T_S$).

For Ni the loading of bulk with carbon atoms is usually performed at temperatures above $1170$\,K~\cite{Odahara:2011iw}. Cooling of the loaded sample leads to the segregation of the carbon atoms at the surface and, depending on the segregation temperature and cooling rate, graphene of different thickness is grown. For single crystalline Ni(111) the growth of single- and double-layer graphene was observed by means of AFM and Raman spectroscopy~\cite{Zhang:2010}. Similar studies performed by means of LEEM [Fig.~\ref{gr-met_segregation}(A)] shows that the first graphene monolayer is formed at $T_S=1125$\,K and the second layer grow at $T_S=1050$\,K~\cite{Odahara:2011iw}. It was also found that the third layer starts to segregate before second layer is formed. Compared to the single crystalline Ni surface, the segregation of carbon on polycrystalline films leads also to the growth of multilayer graphene which is attributed to the presence of grain boundaries in Ni that can serve as nucleation sites for multilayer growth~\cite{Zhang:2010}.

The detailed mechanism of the graphene segregation on metallic surfaces (Ni, as an example) was presented in Ref.~\cite{Liu:2011bl} and it consists of the following steps: (i) dynamic diffusion of the bulk dissolved carbon atoms with moving some of them to the surface, (ii) trapping of the surface-diffusing carbon atoms by defects or step edges on metal surface and creation of the nuclei for the graphene growth, (iii) further graphene nucleation around the graphene centres [Fig.~\ref{gr-met_segregation}(B,a-b)]. The initial result of such process is shown in Fig.~\ref{gr-met_segregation}(B,c), where the start of nucleation of graphene is shown at several places. Decrease of the temperature leads to the rapid growth of graphene islands which finally form the complete graphene layers, that, as seen from the potential energy plot, is energetically favourable [Fig.~\ref{gr-met_segregation}(B,a)]. In such segregation process graphene is formed across the terraces, which are always present on surfaces of metals [Fig.~\ref{gr-met_segregation}(B,d-e)].

The validity of the above presented mechanism of segregation was also demonstrated for graphene growth on single-crystalline Ru(0001)~\cite{Sutter:2008,McCarty:2009fv}. Here initially the carbon atoms were absorbed into the bulk Ru at $1150^\circ$\,C and cooling of the sample to $825^\circ$\,C leads to the segregation of the graphene on the Ru(0001) surface [Fig.~\ref{gr-met_segregation}(C)]. It was found that graphene islands grow via attaching of carbon atoms at the edges and, interestingly, it was found that graphene domains grow parallel to the substrate steps and across steps in the ``downhill'' direction.

\subsection{CVD}

Decomposition of hydrocarbons at high temperatures at the surface of metals is the easiest way to prepare continues graphene layers. This method applied to the polycrystalline metals, like Ni or Cu, demonstrated high perspectives to be used in graphene technology because huge high quality graphene layers of different thicknesses with the size of up to 30 inches were successfully synthesised~\cite{Kim:2009a,Li:2009,Bae:2010}. Several factors determine the growth of graphene on polycrystalline surfaces as pretreatment of the surface, density of the nucleation centers, flow of the hydrocarbon gases, use of additional flow gases (Ar, H$_2$), etc.

\begin{figure}
\includegraphics[width=\linewidth]{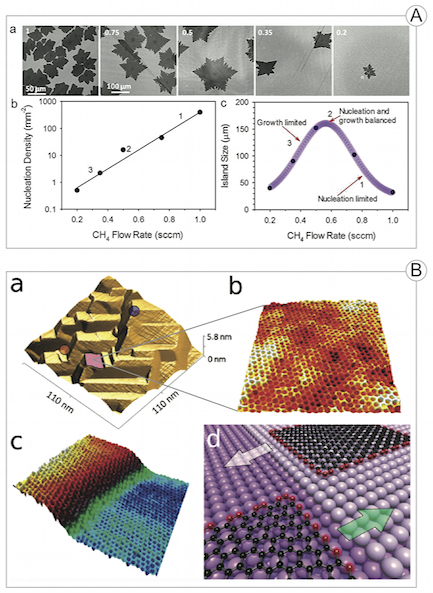}
\caption{(A) Upper row: SEM images of graphene islands on polycrystalline Cu. Numbers denote the flow rate of CH$_4$. Lower row: the corresponding nucleation density and island size as a function of the CH$_4$ flow rate. Data are reproduced from \cite{Eres:2014bc} with permission. (B) STM images of (a) a large area of graphene on polycrystalline Cu, (b) flat area marked in (a) demonstrating atomic resolution of a graphene layer, and (c) a continuous graphene layer covering a monoatomic step of Cu. (d) Schematic presentation of graphene growth on Cu polycrystalline surface. Data are reproduced from \cite{Rasool:2011gx} with permission.}
\label{gr-met_CVD_1}
\end{figure}

Fig.~\ref{gr-met_CVD_1} (A, upper row) shows secondary electron microscopy (SEM) images of graphene grown on unoxidized Cu foil at $1040^\circ$\,C for $0.5$\,hour and different flow rate of CH$_4$~\cite{Eres:2014bc}. The lower row demonstrate the dependence of the nucleation density as well as size of the graphene islands as a function of flow rate showing the optimal conditions for the graphene synthesis where nucleation rate and size of the graphene islands are balanced. Earlier it was shown that for the graphene grown on polycrystalline Cu two predominant planes were observed after graphene growth -- Cu(100) and Cu(311)~\cite{Rasool:2011gx}. In this case graphene overgrow on these surfaces and facets and forms the complete carpet-like layer [Fig.~\ref{gr-met_CVD_1} (B,a-c)]. However, the further experiments with the oxidation and subsequent reduction of Cu foils shows that predominant Cu(100) orientation might be obtained that can be used for the synthesis of the high-quality graphene layers~\cite{Eres:2014bc}.

The shape and size of graphene islands, which later coalescence in the complete layer, also depend on the ratio of H$_2$ or Ar to CH$_4$~\cite{Jacobberger:2013kv,BinWu:2013kw}. As was found in the recent systematic study, the higher ratios of Ar to H$_2$ result in graphene grains with a symmetric dendritic structures, whereas if ratio of Ar to H$_2$ is decreased, the more compact structures of hexagonal or circular shapes were observed~\cite{BinWu:2013kw}. 

The graphene growth on single-crystalline metallic surfaces is performed in UHV conditions that requires their careful preparation via cycles of ion-sputtering (Ne$^+$, Ar$^+$) and annealing at high temperatures~\cite{Coraux:2009,Sicot:2010,Wang:2010ky,Voloshina:2012a,Jacobson:2012gv,Chen:2012gl,Dong:2013jq,Voloshina:2013dq}. In several graphene/metal studies, when carbon has high solubility in the corresponding metal (Ni, Ru, or Rh), thick highly ordered films which were obtained by e-beam evaporation of the respective material were used~\cite{Dedkov:2008d,Dedkov:2008e,Dedkov:2008a,Grueneis:2009,Sutter:2009b,Muller:2009,Dedkov:2010jh,Muller:2011hy,Zeller:2012ha,Dahal:2013do}.

\begin{figure*}
\includegraphics[width=\linewidth]{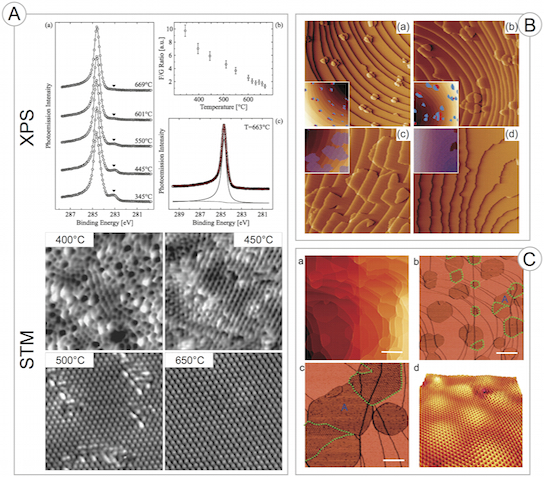}
\caption{(A) Upper panel: (a) C $1s$ XPS spectra of graphene grown at different temperatures on Ni(111), (b) ratio of intensities of two peaks corresponding to C$_3$H$_6$ fragments (F) and main graphene peak (G), (c) Doniach-Sunjic lineshape analysis of the C $1s$ peak of the best-quality graphene. Data are taken from \cite{Grueneis:2009} with permission. Lower panel: STM images of graphene synthesised at different temperatures on Ni(111). Data are taken from \cite{Jacobson:2012gv} with permission. (B) STM images of graphene layers synthesised on Ir(111) after decomposition of C$_2$H$_4$ (partial pressure $5\times10^{-10}$\,mbar) at $1120$\,K for (a) 20\,s, (b) 40\,s, (c) 160\,s, and (d) 320\,s. Data are taken from \cite{Coraux:2009} with permission. (C) STM results for graphene/Cu(111): (a-c) large scale and (d) atomically-resolved small scale images. Data are taken from \cite{Gao:2010a} with permission.}
\label{gr-met_CVD_2}
\end{figure*}

The close-packed surfaces Ni(111) and Co(0001) have lattice constants which are very close to that of graphene (difference is of the order of 1\%) which make them very popular for graphene preparation and studies. Fig.~\ref{gr-met_CVD_2} (A, upper part)~\cite{Grueneis:2009} shows a series of XPS C\,$1s$ spectra of graphene synthesised from C$_3$H$_6$ on Ni(111) at the respective temperatures (marked in the figure) demonstrating the optimal heating conditions: $T\geq650^\circ$\,C. At these parameters the intensity of the peak corresponding to the C$_3$H$_6$ fragments (marked by the black triangle at binding energy of $283$\,eV) vanishes completely. These results were later confirmed in STM studies, which results are compiled in Fig.~\ref{gr-met_CVD_2} (A, lower part)~\cite{Jacobson:2012gv}, showing the high quality graphene grown on Ni(111) at $650^\circ$\,C.

In Refs.~\cite{Weatherup:2011jb,Patera:2013dm} the XPS peak at $283$\,eV was assigned to the carbon surface atoms located at step edges or other defect sites of the Ni(111) surface (this peak also appears if the hydrocarbon source is replaced the solid carbon source). 

Later XPS and LEED experiments performed under \textit{in operando} conditions~\cite{Patera:2013dm} show that if clean Ni(111) surface is exposed to hydrocarbons below $500^\circ$\,C, then it leads to the formation of  Ni$_2$C which later converts to monolayer graphene via an in-plane mechanism~\cite{Lahiri:2011iu}. Above $500^\circ$\,C graphene grows via replacement of Ni surface atoms leading to embedded epitaxial and/or rotated graphene domains~\cite{Patera:2013dm}.

The studies of graphene growth on Ru(0001) demonstrate that faceting of the metallic surface occurs with formation of large terraces where graphene forms a perfectly ordered layer~\cite{Loginova:2009ug,Gunther:2011jq}. Graphene in this system overgrows the metal steps and the process of growth is governed by the big Ru-atoms transport and these atoms migrate from ``etched'' terraces underneath a graphene layer~\cite{Gunther:2011jq}.

Surprisingly, the studies of the graphene CVD growth on Rh(111) showed that the temperature window for this process is very narrow: graphene was only found on the surface in the range between $1016$\,K and $1053$\,K~\cite{Dong:2012hq,Dong:2013jq,Wang:2010ky,Stojanov:2014fk}. Below this temperature range a rhodium carbide is formed. These results were compared with those obtained during temperature programmed growth (TPG) procedure (C$_2$H$_4$ is adsorbed at room temperature and then metal heated to high temperature) and it was found that in this case the temperature range for the graphene formation is much broader -- between $808$\,K and $1053$\,K~\cite{Dong:2012hq}. 

Formation of the graphene/Ir(111) system was intensively studied in a series of LEEM and STM experiments~\cite{Coraux:2008,Coraux:2009,vanGastel:2009,NDiaye:2009,Hattab:2012fq}. Fig.~\ref{gr-met_CVD_2}(B) shows a series of STM images collected after exposure of Ir(111) to C$_2$H$_4$ at $1120$\,K for different time. Graphene coverage grows linearly in the beginning, asymptotically approaching full layer for the longer time. Compared to graphene/Ru(0001), on Ir(111) a graphene layer grows in both directions -- ``uphill'' and ``downhill'' across the metallic steps with preferential nucleation around the step edges, but growth on the flat terraces was also detected. Contrary to many other transitional metals, the growth of well-ordered graphene layer on Ir(111) can be performed in very wide temperature range of $970$\,K - $1470$\,K~\cite{Coraux:2009}.

Compared to $d$-metals with open valence band shells, the CVD growth of graphene on the close $d$-shell metals is quite difficult. For example CVD graphene-growth on Cu(111) is possible only at temperatures very close to the melting temperature. Fig.~\ref{gr-met_CVD_2}(C) shows the STM results of such studies~\cite{Gao:2010a}. Here a graphene growth on Cu(111) was performed at very high partial pressure of C$_2$H$_4$ ($10^{-5}$\,mbar and dosing via nozzle) and cycles of fast sample flash-annealing to $1000^\circ$\,C, which is very close to the melting point of Cu ($1083^\circ$\,C). In these experiments graphene layers of $0.35-0.8$ monolayer were studied. Later, the problem of the usage of the high temperature for graphene synthesis, which is very close to the melting point, was solved via irradiation of Cu(111) and Au(111) with ethylene ($500$\,eV at $800^\circ$\,C) followed by the additional annealing at $900-950^\circ$\,C~\cite{MartinezGalera:2011cx}. 

\subsection{MBE growth}

\begin{figure}
\includegraphics[width=\linewidth]{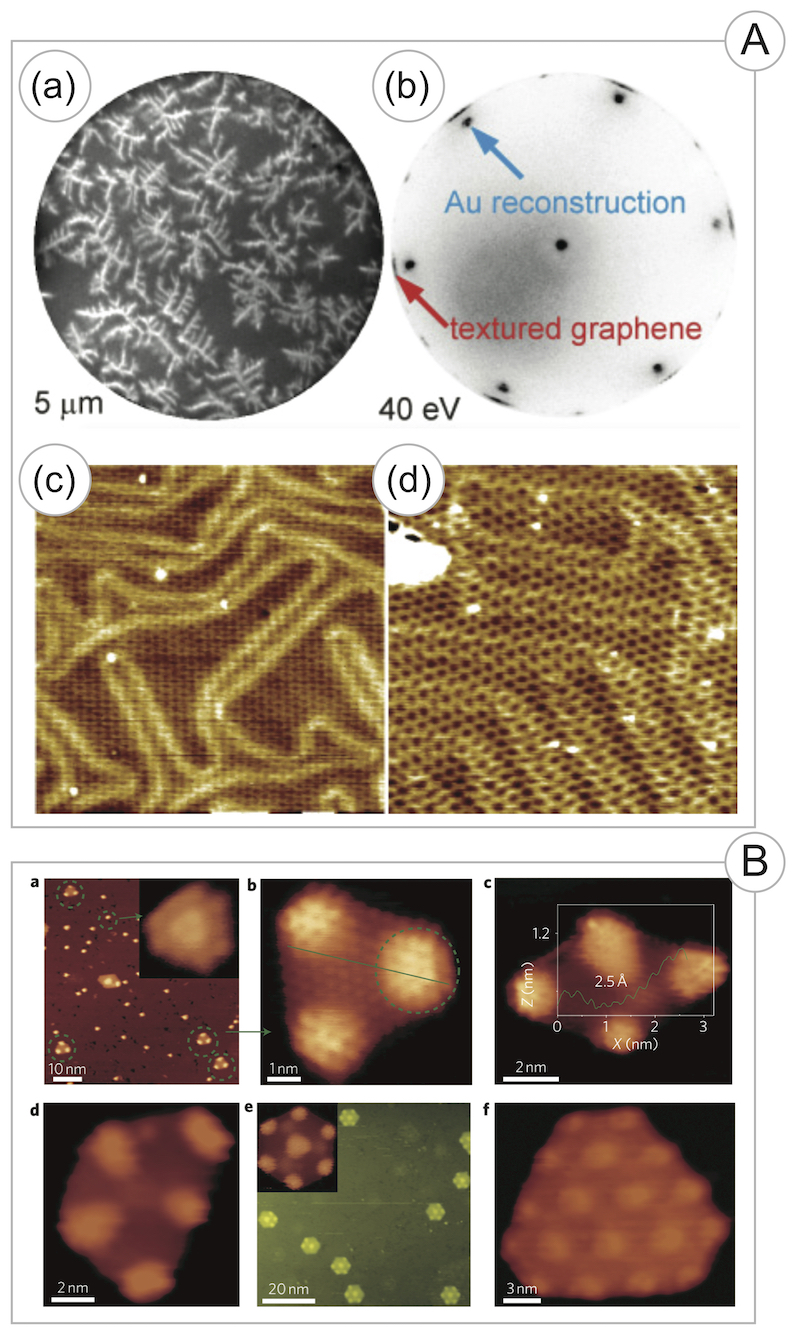}
\caption{(A) (a) LEEM image (field of view: $5$\,$\mu$m) of graphene/Au(111) grown by MBE at $950^\circ$\,C. The bright features are dendritic islands of graphene. The grey background is the bare Au surface. (b) LEED (40 eV) from an area of 2 $\mu$m in diameter. (c) STM images of graphene moir\'e modified by the Au herringbone. Data are taken from \cite{Nie:2012jd} with permission. (B) STM images of graphene quantum dots on Ru(0001) synthesised from the pre-deposited C$_{60}$ molecules. Data are taken from \cite{Lu:2011bg} with permission.}
\label{gr-met_MBE}
\end{figure}

This method is usually used for the preparation of the epitaxial graphene layers on the insulating and semiconducting surfaces at high temperatures~\cite{Zhou:2012cp,Lippert:2013vp,Lippert:2014va} as it was suggested that it can be easily adopted for the mass production and further graphene technologies. Recently it was shown that MBE growth can be successfully used for the preparation of graphene on Au(111) in UHV conditions~\cite{Nie:2012jd,Wofford:2012fp} [Fig.~\ref{gr-met_MBE}(A)]. Carbon was deposited from an electron-beam-heated graphite rod at substrate temperatures ranging from $770^\circ$\,C to $940^\circ$\,C. The quality of graphene/Au(111) was monitored \textit{in situ} by means of LEEM/$\mu$LEED and STM. On the initial stage of growth, graphene islands have a dendritic structure and, interestingly, in STM it was possible to resolve the untouched herringbone reconstruction of Au(111) that indicates extremely small attraction energy between graphene and Au ($<13$\,meV/C-atom)~\cite{Nie:2012jd}. MBE growth of graphene of different thickness (up to $3.5$ monolayers) was also demonstrated on Ni(111)/MgO(111)~\cite{Wofford:2014eg}.

Deposited carbon-contained molecules on the surface of metals might be considered as a solid source of carbon atoms for the synthesis of graphene~\cite{Lu:2011bg,Chen:2012gl}. For example, it was shown that C$_{60}$ can be used for the preparation of well-ordered graphene quantum dots of limited size on Ru(0001) using the high catalytic activity of this surface~\cite{Lu:2011bg} [Fig.~\ref{gr-met_MBE}(B)]. If two precursors for the graphene growth are compared, C$_2$H$_4$ and C$_{60}$, then it was found that for the former one graphene is predominantly formed at the steps of substrate, whereas for the later case graphene dots have very low mobility at high temperature and might be randomly distributed over the Ru surface.

\section{Crystallographic structure of graphene on metals}

\begin{figure}
\includegraphics[width=\linewidth]{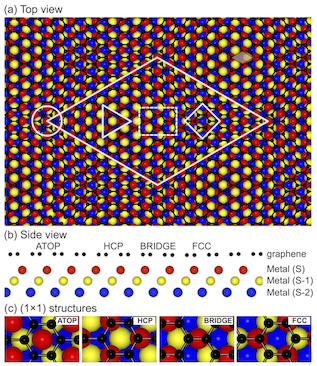}
\caption{Top (a) and side (b) views of a graphene layer on the close-packed (111) metal surface: $(n\times n)$graphene/$(m\times m)$Metal(111) (here: $n=10$, $m=9$). High symmetry adsorptions sites of the Metal(111) surfaces are marked: circle (ATOP), triangle (HCP), rhombus (FCC), rectangle (BRIDGE). (c) Local high-symmetry $(1\times 1)$ structures of the graphene/Metal(111) interface.}
\label{gr-met-structures}
\end{figure}

Synthesis of high-quality graphene layers on metals requires the perfect experimental conditions and always involves a high temperature. Any surface imperfections of metal or slight change of the synthesis conditions might lead to the appearence of different defects in graphene (vacancies, defect lines, etc.) as well as to the different alignments of a graphene lattice on the surface of metal. Here we consider situations of the ideal alignment of graphene and metal close-packed surfaces when lattice vectors of both sublattices are parallel to each other, the so-called $R0^\circ$ structures (Fig.~\ref{gr-met-structures}). The discussion of other possible geometrical structures of graphene on metals as well as their interpretations can be found elsewhere~\cite{Coraux:2008,Loginova:2009,Gao:2010a,Gao:2011,Man:2011bo,Starodub:2011a,Meng:2012dr,Hermann:2012dy,Jeon:2013ek,Dedkov:2014di}.

Generally there is a lattice mismatch between graphene and metallic surfaces which is ranged from $1.3$\% for Ni(111) to, e.\,g., $12.8$\% for Pt(111) [Fig.~\ref{gr-met-structures}(a,b)]. Therefore during formation of graphene on Ni(111) or Co(0001), graphene adopts the lattice constant of the metallic substrate and the so-called $(1\times1)$ structures are formed [Fig.~\ref{gr-met-structures}(a-c)] (although, as stated earlier, the slight variation of the synthesis parameters might lead to the formation of rotated structures, which were observed in the experiment~\cite{Dedkov:2010jh,Dahal:2012jh,Jacobson:2012be}). For $4d$ and $5d$ metals adsorption of graphene on their close-packed surfaces always leads to the formation of the moir\'e structures, which orientation depends on the relative orientation of the lattice vectors of graphene and Metal(111) surface, whereas the corrugation of such structure is defined by the interplay between geometry and the local graphene/metal interaction around the high symmetry positions of the moir\'e structure.

Historically, the first method which gave information about crystallographic structure was LEED~\cite{Hagstrom:1965vh,Lyon:1967,May:1969uj,Grant:1970}. Presently this method is used for the qualitative characterisation as well as for quantitative $I-V$ LEED analysis of the graphene/metal structures~\cite{Gamo:1997,Moritz:2010,Parreiras:2014in}. However, the main method which provides an accurate analysis of the crystallographic structure on the large and small scale is STM, giving also access to the local electronic structure.

\begin{figure}[b]
\includegraphics[width=\linewidth]{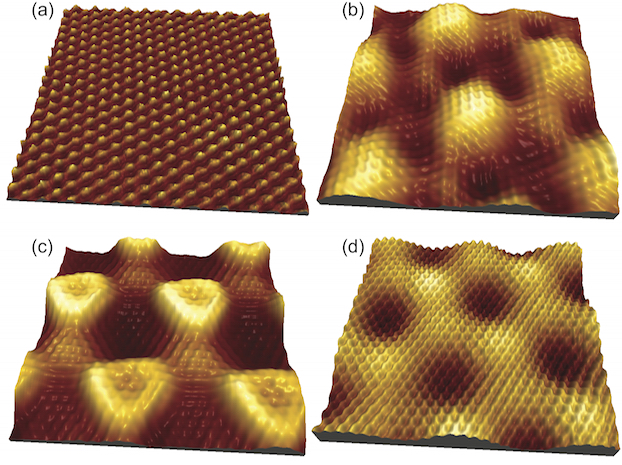}
\caption{STM images of (a) gr/Ni(111), (b) gr/Rh(111), (c) gr/Ru(0001), and (d) gr/Ir(111).}
\label{gr-met_STM}
\end{figure}

Fig.~\ref{gr-met_STM} shows a representative STM images of (a) $(1\times1)$gr/$(1\times1)$Ni(111), (b) $(12\times12)$gr/$(11\times11)$Rh(111), (c) $(13\times13)$gr/$(12\times12)$Ru(0001), and (d) $(10\times10)$gr/$(9\times9)$Ir(111). 
 
 \begin{figure*}[t]
\includegraphics[width=\linewidth]{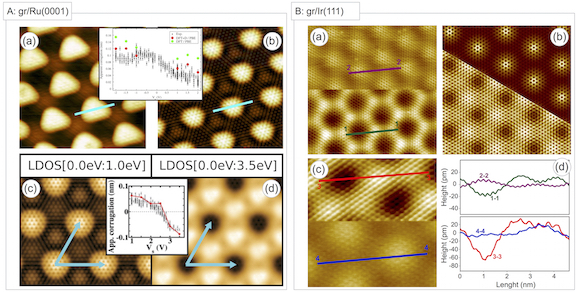}
\caption{(A) STM images of gr/Ru(0001): (a) and (b) are experimental and simulated STM images, respectively, obtained for the bias voltage $U_T=-1$\,V; (c,d) Simulated STM images for $U_T=+1$\,V and $U_T=+3.5$\,V. Insets show the apparent corrugation of graphene as a function of the bias voltage (experimental and theoretical data). Data are taken from \cite{Stradi:2011be,Stradi:2012hw} with permission. (B) Combined STM/AFM results for gr/Ir(111): (a) and (b) are experimental and simulated STM images, respectively, where lower part of every image corresponds to $U_T=-0.6$\,V and upper one to $U_T=-1.8$\,V. (c) STM (top) and AFM (bottom) of gr/Ir(111) where scanning mode was changed \textit{on-the-fly} in the middle of the scan. Scanning parameters: $U_T=0.03$\,V, $I_T=1$\,nA, $\Delta f=-475$\,mHz. (d) Extracted height profiles from STM and AFM data (see a-c).}
\label{STM_grRu_grIr}
\end{figure*}
 
Graphene on Ni(111) has a $(1\times1)$ structural periodicity as the lattice constants of graphene and Ni(111) are very close to each other [Fig.~\ref{gr-met_STM}(a)]. Structural measurements based on the $I-V$ LEED and photoelectron diffraction (PED) analysis give HCP ($top-fcc$) structure as the most stable arrangement of carbon atoms on Ni(111) [Fig.~\ref{gr-met-structures}(c)], with the distance between graphene and top Ni layer of $2.11$\,\AA\ and very small corrugation of graphene of $0.05$\,\AA~\cite{Gamo:1997,Parreiras:2014in}. These results were analysed in many DFT calculations (LDA, PBE, PBE-D, PBE-D2) and very good agreement was found between experiment and theoretical data~\cite{Bertoni:2004,Weser:2011,Voloshina:2011NJP,Dzemiantsova:2011bv,Adamska:2012,Voloshina:2013cw,Voloshina:2014iy,Dahal:2014jv}. The state-of-the-art calculations at the PBE-D2 level give the maximal bonding energy between graphene and Ni(111) of about $160$\,meV/C-atom~\cite{Voloshina:2011NJP,Adamska:2012}, which is much smaller than the lower limit for the chemical bonds, placing this value in the range of the van der Waals (vdW) interactions. In this system, graphene is strongly $n$-doped via electron transfer from the Ni\,$4s$ on the graphene\,$\pi^*$ states, that leads to the space-, energy-, and wave-vector-overlap of Ni\,$3d$ and graphene\,$\pi$ states. As a result, several so-called interface states are formed, which are composed from Ni\,$3d_{xz,yz,z^2}$ and C\,$p_z$ orbitals (see also discussion below)~\cite{Voloshina:2012c,Voloshina:2014jl}. Similar situation is also realised for the graphene/Co(0001) interface~\cite{Eom:2009,Voloshina:2011wa}.

Adsorption of graphene on other metals, like Cu, $4d$ or $5d$ metals, always leads to the formation of the moir\'e structures [see Figs.~\ref{gr-met_CVD_2}(C), \ref{gr-met_MBE}, \ref{gr-met_STM}, \ref{STM_grRu_grIr}]. Below we consider two interesting cases of graphene on Ru(0001) and Ir(111), where correspondingly strongly and weakly corrugated graphene layers are formed.

Graphene/Ru(0001) was investigated by different experimental methods, which give structural information (corrugation of a graphene layer): $I-V$ LEED ($1.5$\,\AA)~\cite{Moritz:2010}, surface x-ray diffraction (SXRD) ($0.82$\,\AA)~\cite{Martoccia:2008,Martoccia:2010}, STM (between $0.5$\,\AA\ and $1.1$\,\AA\ depending on the tunneling conditions)~\cite{Marchini:2007,VazquezDeParga:2008,Borca:2009,Sutter:2009hy,Borca:2010,Stradi:2011be,Stradi:2012hw}. 

Presently, the widely accepted structure of graphene/Ru(0001) is $(25\times25)$gr/$(23\times23)$Ru(0001), meaning that $25$ unit cells of graphene are aligned along $23$ unit cells of the Ru(0001) surface~\cite{Martoccia:2008,Moritz:2010,Martoccia:2010} (see Refs.~\cite{Wang:2008,Jiang:2009,Brugger:2009,Wang:2010jw,Altenburg:2010ke,Wang:2011hh,Alfe:2013im} for the discussion of other smaller size models). These results were obtained in $I-V$ LEED and SXRD experiments and compared with the DFT calculations within the GGA approach (PBE functional without inclusion of vdW interactions)~\cite{Moritz:2010}. This modelling gives a corrugation of graphene of $1.59$\,\AA\ (vs. $1.53\pm0.2$\,\AA\ obtained in experiment) and correctly reproduce the obtained experimental data ($I-V$ LEED curves, corrugation of the graphene layer, variation of bonds in the graphene layer, and corrugation of the topmost Ru layer).

However, the discussed model is quite expensive with respect to the computational time, and another model, namely $(11\times11)$gr/$(10\times10)$Ru(0001), was used for the investigation of the effect of vdW interactions on the electronic structure of gr/Ru(0001)~\cite{Stradi:2011be,Stradi:2012hw}. In these works the PBE-D2 approach proposed by Grimme which includes vdW corrections~\cite{Grimme:2006} was used. This modelling shows that the corrugation of a graphene layer is $1.195$\,\AA\ (minimal distance between graphene and the top Ru layer is $2.195$\,\AA)~\cite{Stradi:2011be}, which is by $0.40-0.55$\,\AA\ smaller (depending on the structural model) compared to the values obtained without the vdW interaction~\cite{Wang:2008,Jiang:2009,Moritz:2010}. Inclusion of the vdW interaction also leads to the dramatic increase of the bonding energy from $27$\,meV/C-atom (no vdW) to $206$\,meV/C-atom (however as mentioned above this value is below the lower limit for the estimation of the covalent chemical bonding strength). The presented model correctly reproduces the STM data for gr/Ru(0001) [Fig.~\ref{STM_grRu_grIr}(A)]~\cite{Stradi:2011be,Stradi:2012hw}: STM topography as well as a value of corrugation (insets). For example, the inversion of the imaging contrast in STM was reproduced for the bias voltages around $+2.5$\,V: above this point the topographically highest ATOP places of the gr/Ru(0001) structure are imaged as a dark areas in the STM data [Fig.~\ref{STM_grRu_grIr}(A,c-d)]. This effect was explained by the formation of the interface states at the HCP and FCC areas as a result of overlap of an unoccupied Ru(0001) surfaces resonance with the first image state component localized at these places of the moir\'e structure~\cite{Borca:2010jj,Stradi:2012hw}.

Graphene on Ir(111) forms a moir\'e structure with a periodicity of $(10\times10)/(9\times9)$ [Fig.~\ref{gr-met_STM}(c)]. If low bias voltages between sample and tip are used for the imaging of this system (below $\pm1$\,V, which is typical for STM experiments on metallic systems), then the \textit{inverted} contrast in STM is observed, when topographically highest ATOP places are images as dark areas [Figs.~\ref{gr-met_STM}(c) and \ref{STM_grRu_grIr}(B,a)]~\cite{NDiaye:2008qq,Voloshina:2013dq}. Increasing the bias voltage leads to the inversion of the imaging contrast and it becomes a \textit{direct} one [Fig.~\ref{STM_grRu_grIr}(B,a)]. Similar to gr/Ru(0001), STM experiments on gr/Ir(111) also show the bias dependence of the corrugation [Fig.~\ref{STM_grRu_grIr}(B,d)], which can vary between $0.15$\,\AA\ and $1.0$\,\AA. Careful combined STM/AFM experiments of gr/Ir(111)~\cite{Voloshina:2013dq,Dedkov:2014di} trace the changes in the imaging contrast and allow the accurate identification of the high-symmetry positions in this structure [Fig.~\ref{STM_grRu_grIr}(B,c)]: ATOP positions are imaged as bright areas in the attractive regime ($\Delta f<0$) of NC-AFM. The obtained corrugation of $0.3$\,\AA\ of the graphene layer in AFM experiment witht W-tip is very close to the theoretically calculated value (see discussion below). If CO-terminated tip is used, then corrugation of $0.42-0.56$\,\AA\ was obtained supported by the parallel $I-V$ LEED measurements~\cite{Hamalainen:2013jj}. Contrary to these results, the much larger corrugation of $0.6$\,\AA\ and $1.0$\,\AA\ for $0.39$\,ML and $0.63$\,ML graphene on Ir(111), respectively, was obtained from x-ray standing wave (XSW) experiments~\cite{Busse:2011} that was assigned to the increasing of the number of wrinkles on the surface. 

\begin{figure*}[t]
\includegraphics[width=\linewidth]{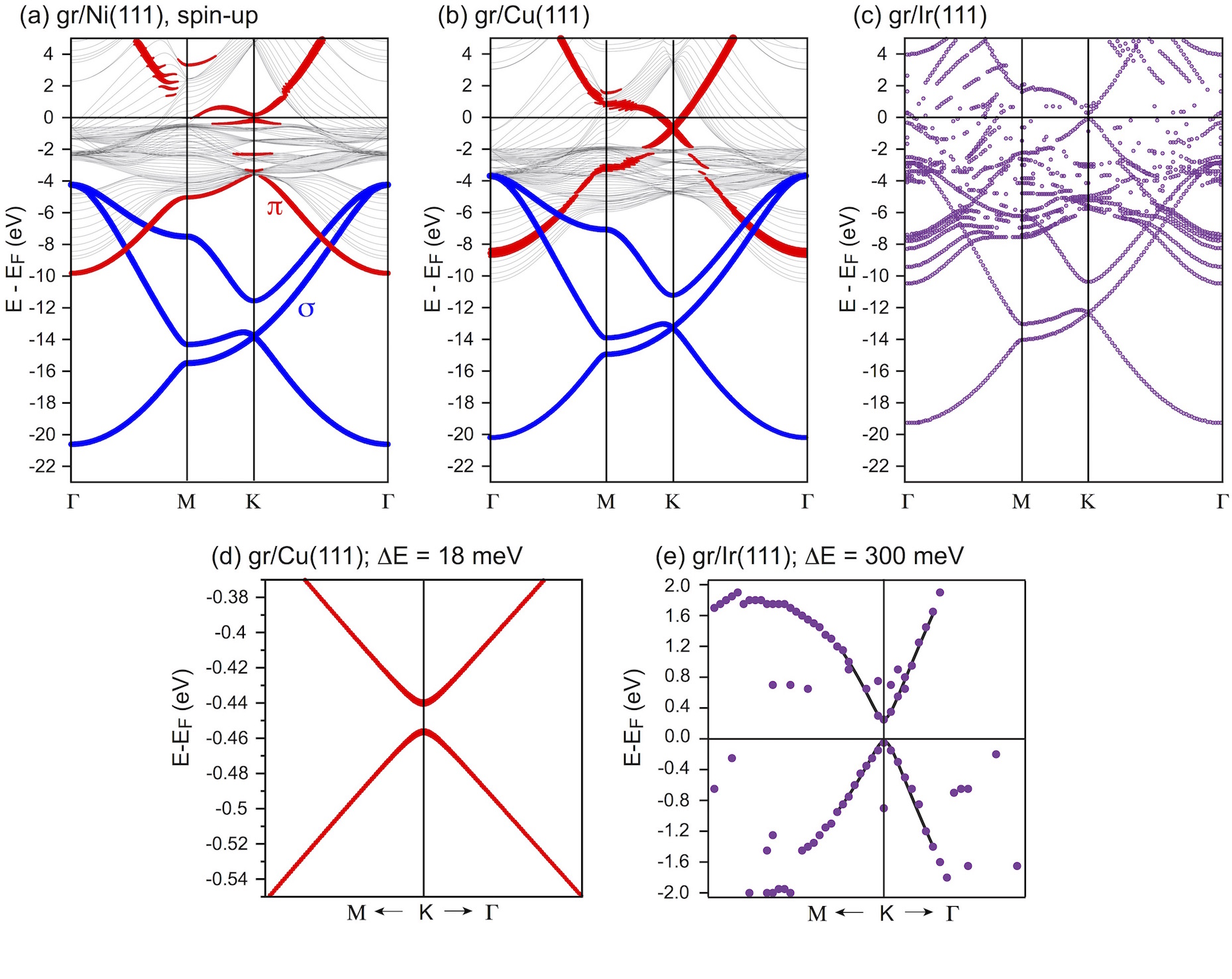}
\caption{Calculated band structures of (a) graphene/Ni(111) (spin-up) and (b) graphene/Cu(111) along high-symmetry directions of the graphene Brillouine zone. Calculations are performed for $(1\times1)$ structures (small lattice mismatches are ignored in both cases). The $\pi$ and $\sigma$ graphene-derived bands are highlighted by the red and blue colours, respectively. (c) Band structure of $(10\times10)$graphene/$(9\times9)$Ir(111) unfolded for the graphene $(1\times1)$ primitive cell. Panels (d) and (e) show the zoomed regions of the respective electronic structures where energy gap for the graphene $\pi$ states is open. The corresponding energy gaps are marked in the figures.}
\label{BANDS_grNi_grCu}
\end{figure*}

The graphene/Ir(111) system was modelled within the framework of the DFT theory with vdW interactions accounted by means of a semiempirical DFT-D2 approach proposed by Grimme~\cite{Grimme:2006} or using a non-local correlation functional vdW-DF proposed by Dion \textit{et al.}~\cite{Dion:2004,Klimes:2009ei,Klimes:2011vk} These calculations lead to the following values for the graphene corrugation and the mean distance between the graphene layer and the top-most Ir layer: $0.31$\,\AA\ and $3.39$\,\AA~\cite{Voloshina:2013dq}, $0.37$\,\AA\ and $3.28$\,\AA~\cite{Pacile:2013jc}, $0.35$\,\AA\ and $3.41$\,\AA~\cite{Busse:2011}. (The previous calculations performed at the GGA level without vdW interactions included give the minimal and maximal distances between graphene and Ir(111) of $3.77$\,\AA\ and $4.04$\,\AA, respectively~\cite{Ndiaye:2006}.) All these simulations predict the relatively large distance between graphene and Ir(111), which is very close to the distance between carbon layers in graphite. Analysis of the interaction shows that GGA approach leads to repulsion between graphene and Ir(111) ($\approx+20$\,meV/C-atom) and only inclusion of the vdW interaction gives the attraction in this system~\cite{Busse:2011}. It was found that here graphene is slightly $p$-doped and there is a local overlap of the Ir\,$5d_{z^2}$ and C\,$p_z$ orbitals at HCP and FCC high-symmetry places of the graphene/Ir(111) structure that leads to the appearence of the so-called hybrid states in the valence band of this system~\cite{Busse:2011,Voloshina:2013dq}. Existence of such states was used for the explanation of the bias dependence of the change of the imaging contrast in STM data: ATOP positions are imaged as dark places at low bias voltages and become bright with increasing the voltage~\cite{Voloshina:2013dq}. These results point the importance of the graphene-metal interaction in this system~\cite{Voloshina:2013dq,Starodub:2011a}, although in the beginning it was believed that graphene is nearly free-standing on Ir(111)~\cite{Pletikosic:2009}.

The relatively weak interaction between a graphene layer and Ir(111) as well as between graphene and Pt(111) leads to the fact that slight variation of the graphene synthesis parameters (temperature or/and partial pressure of hydrocarbons) might cause the appearance of the rotational domains of graphene on the metallic surfaces, where graphene lattices are misoriented with respect to each other~\cite{Coraux:2008,Loginova:2009,Gao:2010a,Gao:2011,Man:2011bo,Starodub:2011a,Meng:2012dr,Hermann:2012dy,Jeon:2013ek,Dedkov:2014di}.

\section{Electronic structure of graphene on metals}\label{ELSTR}

Adsorption of a graphene layer on metallic surface always leads to the modification of the electronic structure of graphene. According to the modern state-of-the-art electronic structure calculations graphene is always weakly bonded to metals: the bonding energy at the interface was found in the range of $50-200$\,meV/C-atom (see examples above), which is much lower than the lowest limit which is used for the description of chemical bonds. However, in all cases the original band structure of graphene around $E_F$ (linear dispersion and the Dirac cone) is distorted via doping ($n$ or $p$) or/and via overlap with the valence band states of metal. The progress in the complete understanding of all changes in the electronic structure of graphene upon its adsorption on metals was reached in the recent publications~\cite{Wintterlin:2009,Voloshina:2012c,Voloshina:2014jl}. The proposed \textit{universal} model for the description of the graphene-metal interaction devides this process into two steps: (i) doping of graphene, $n$ or $p$, by mobile $s$-electrons, that increases or decreases, respectively, the strength of the vdW interaction between graphene and metal and might lead, in the case of the $n$-doping, to the (ii) effective space-, energy-, and wave-vector-overlapping of the C\,$p_z$ states of graphene and metallic $d$ states. Depending on the relative energy-overlap of the $\pi$ and $d$ bands the Dirac cone in the electronic structure of graphene is fully destroyed (case of graphene/Ni) or symmetry energy gap is opened directly at the Dirac point~\cite{Voloshina:2012c,Voloshina:2014jl}.

\begin{figure*}[t]
\includegraphics[width=\linewidth]{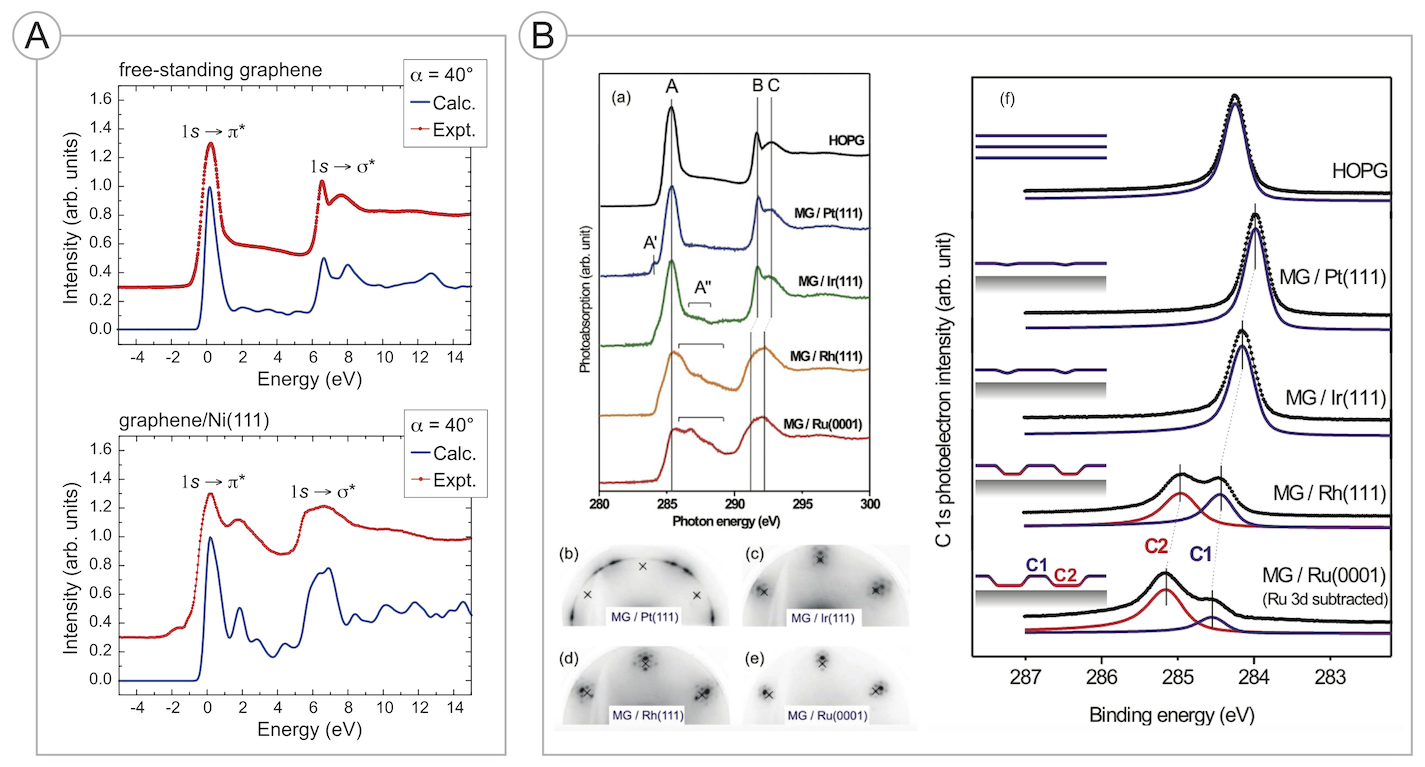}
\caption{A: comparison of the experimental and calculated NEXAFS spectra for (a) graphene and (b) graphene/Ni(111) ($\alpha$ is the incident angle of the linearly polarized light on the surface). Data are taken from ~\cite{Voloshina:2013cw} with permission. B: (a) C\,$K$-edge NEXAFS spectra, (b-e) LEED images, and (f) C\,$1s$ XPS spectra of different graphene moir\'e structures on $4d$ and $5d$ metals. Data are taken from \cite{Preobrajenski:2008} with permission.}
\label{NEXAFS_gr-met}
\end{figure*}

Fig.~\ref{BANDS_grNi_grCu} shows calculated band structures of (a) graphene/Ni(111) (spin-up) and (b,d) graphene/Cu(111) for the HCP ($top-fcc$) arrangement (small lattice-mismatch is ignored) [Fig.~\ref{gr-met-structures}(c)]. The bonding energies and the distances between graphene and metal surfaces are $67-160$\,meV/C-atom and $2.11$\,\AA, respectively, for gr/Ni(111) (depending on the used functional)~\cite{Voloshina:2011NJP,Dzemiantsova:2011bv,Mittendorfer:2011,Adamska:2012,Voloshina:2013cw,Voloshina:2014iy,Dahal:2014jv} and $94$\,meV/C-atom and $3.02$\,\AA, respectively, for gr/Cu(111) (PBE-D2). The resulting band structure of these systems is very different. In case of graphene/Ni(111), the initial $n$-doping of graphene shifts the Dirac cone below $E_F$, where energy- and wave-vector-overlap of the graphene\,$\pi$ states with Ni\,$3d$ states, accompanied by the space overlap of the corresponding orbitals, leads to the formation of the so-called \textit{hybrid} states at the interface between graphene and Ni(111). The Dirac cone as well as original electronic structure of graphene in the vicinity of $E_F$ are fully destroyed~\cite{Voloshina:2013cw,Voloshina:2014iy}. If graphene is adsorbed on Cu(111), then in the first step the doping shifts the graphene Dirac cone  below $E_F$, but overlap of the graphene\,$\pi$ and Cu\,$3d$ states appears at the large binding energies below the Dirac cone. The deep analysis of this system shows that the so-called symmetry energy gap is opened directly at the Dirac cone of graphene due to the overlap of the C\,$p_z$ states from two carbon sublattices of graphene with the Cu\,$3d$ states of different symmetry from the top-most Cu layer~\cite{Vita:2014aa,Voloshina:2014jl}.

Fig.~\ref{BANDS_grNi_grCu}(c,e) shows calculated band structure of the $(10\times10)$graphene/$(9\times9)$Ir(111) system unfolded to the graphene $(1\times1)$ primitive unit cell according to the procedure described in Refs.~\cite{Popescu:2012bq,Medeiros:2014ka} with the code \texttt{BandUP}~\cite{Medeiros:2014ka}. Although, from the first look the band structure have a \textit{spaghetti}-style, one can clearly recognise the band dispersions of the graphene-derived $\pi$ and $\sigma$ states. These calculations show that graphene in this system is $p$-doped and the Dirac point is located at $150$\,meV above $E_F$ and the energy gap of $\approx300$\,meV is opened for the $\pi$ states at the $K$ point. This result is in good agreement with recent ARPES data for graphene/Ir(111)~\cite{Starodub:2011a}, where lowering of the Dirac point was induced by the K adsorption that allows observation of the energy gap of $\approx100$\,meV at the $K$-point.

Core-level based electron spectroscopies (XPS and NEXAFS) are powerful methods for investigation of the electronic structure of the graphene-based systems. Fig.~\ref{NEXAFS_gr-met} shows the NEXAFS and XPS spectra of different graphene-metal systems in comparison with those for graphite (consists of the layers of graphene separated by $3.4$\,\AA)~\cite{Preobrajenski:2008,Weser:2010,Rusz:2010,Weser:2011,Voloshina:2011NJP,Voloshina:2013cw}. NEXAFS spectra of graphite are very good example for the demonstration of the \textit{search-light}-like effect as the relative intensities of the $1s\rightarrow\pi^*$ and $1s\rightarrow\sigma^*$ absorption bands are changed if the incident angle $\alpha$ is varied~\cite{Weser:2010}.

\begin{figure*}[t]
\includegraphics[width=\linewidth]{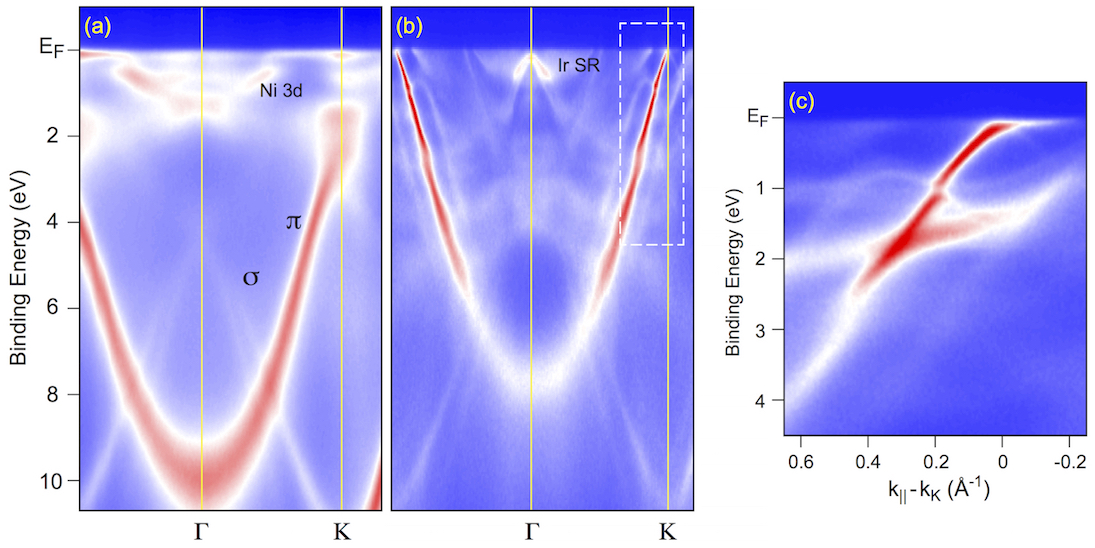}
\caption{ARPES intensity maps for (a) graphene/Ni(111) and (b) graphene/Ir(111) measured along the $\Gamma-K$ direction of the Brillouine zone. (c) Zoom of the ARPES map of graphene/Ir(111) marked by the dashed line in (b). Photon energy is $h\nu=65$\,eV.}
\label{ARPES_grNi_grIr}
\end{figure*}

NEXAFS and C\,$1s$ spectra can be used for the investigation of the orbital overlap of the valence band states of graphene and metallic substrate. In case of graphene on Ni(111), Rh(111), and Ru(0001) the strong modification of the $1s\rightarrow\pi^*$ absorption band is observed (Fig.~\ref{NEXAFS_gr-met}) compared to the one for graphite, indicating the strong gr\,$\pi$ -- Metal\,$d$ overlap at the interface with a formation of \textit{hybrid} states, whereas NEXAFS spectra for gr/Pt(111) and gr/Ir(111) are moderately influenced. [The respective changes in the $1s\rightarrow\sigma^*$ absorption band for gr/Ni(111), gr/Rh(111), and gr/Ru(0001), compared to HOPG and gr/Pt(111), are related to the partial $sp^2$-to-$sp^3$ re-hybridization for the former systems~\cite{Preobrajenski:2008,Dedkov:2010jh}.] C\,$1s$ XPS spectra for these systems are different reflecting the influence of the crystallographic structure of the system on the electronic structure. In case of gr/Ni(111) this spectrum consists of one line (shifted by $\approx0.4$\,eV with respect to the spectrum for graphite due to the charge transfer) indicating that graphene on Ni(111) is flat [see STM image in Fig.~\ref{gr-met_STM}(a)] and that the inequivalency in position of two carbon atoms in the graphene lattice has extremely small influence on the electronic structure. C\,$1s$ spectra for the graphene moir\'e structures are different. Following the previous considerations, the mean distance between graphene and Ir(111) or Pt(111) is quite large  of $\approx3.4$\,\AA\ with a graphene corrugation of about $0.3$\,\AA, and the C\,$1s$ spectra for these systems have only one component. For the gr/Rh(111) and gr/Ru(0001) systems, graphene is strongly corrugated ($\approx1.1$\,\AA) with places where the distance between graphene and metallic surface is close to the one for gr/Ni(111) ($\approx2.1$\,\AA) and where strong orbital overlap occurs. Therefore the C\,$1s$ XPS spectrum for these graphene-moir\'e structures consists of two components corresponding to \textit{valleys} (high binding energy component C2) and \textit{hills} (low binding energy component C1).

Strong orbital intermixing of the graphene\,$\pi$ and Ni\,$3d$ valence band states at the interface, discussed above, leads to the appearence of the induced magnetic moment of carbon atoms which was detected by means of XMCD~\cite{Weser:2010,Weser:2011,Matsumoto:2013eu}. This observation is supported by the electronic structure calculations of the induced magnetic moments~\cite{Bertoni:2004,Weser:2011} as well as by the spin-resolved ARPES measurements of the exchange splitting of the $\pi$ band in the graphene/Ni(111) system~\cite{Dedkov:2010jh}. 

Fig.~\ref{ARPES_grNi_grIr} shows ARPES intensity maps measured for two systems, graphene/Ni(111) (a) and graphene/Ir(111) (b-c), which can be considered as two limit cases of the graphene-metal systems where \textit{strong} and \textit{weak} overlaps, respectively, of the valence band states at the interfaces are observed.

Graphene/Ni(111) interface has $(1\times1)$ structure and strong orbital mixing of the graphene\,$\pi$ and Ni\,$3d$ valence band states is found at the interface~\cite{Bertoni:2004,Gruneis:2008,Dedkov:2010jh,Voloshina:2011NJP} [Figs.~\ref{BANDS_grNi_grCu}(a) and \ref{ARPES_grNi_grIr}(a)]. In ARPES maps one can clearly identify graphene $\pi$ and $\sigma$ states which are shifted to higher binding energies (compared to free-standing graphene) due to the electron doping. The effect of orbital mixing is clearly visible around the $K$-point of the Brillouine zone where several \textit{hybrid} states are formed in the energy range between $E_F$ and $2.3$\,eV. These two effects, doping and states-overlap, shift $\pi$ states by $\approx2.4$\,eV to higher binding energies, compared to $\approx1$\,eV for the $\sigma$ states, which are influenced by the doping only. Valence band states of Ni are also strongly modified as compared to the clean Ni(111) surface do to the charge redistribution at the interface~\cite{Dedkov:2008e}. The similar behaviour of the valence band states is also found in other ARPES studies of the graphene-metal systems, where significant overlap of the valence states occurs at the interface: graphene/Co(0001)~\cite{Varykhalov:2009}, graphene/Rh(111)~\cite{Sicot:2012}, graphene/Ru(0001)~\cite{Sutter:2009,Brugger:2009,Enderlein:2010}, graphene/Re(0001)~\cite{Papagno:2013ew}.

\begin{figure*}[t]
\includegraphics[width=\linewidth]{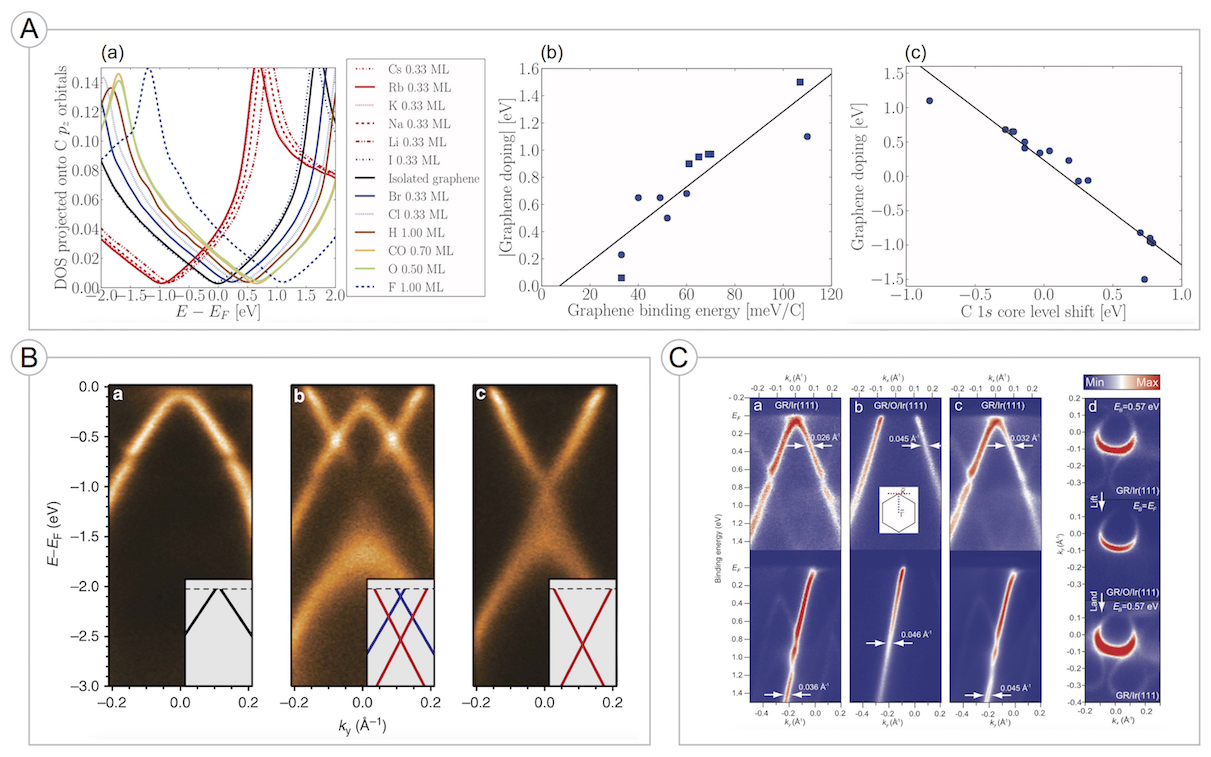}
\caption{A: (a) Partial C\,$p_z$ density of states for different intercalation graphene/X/Ir(111) systems as compared to the free-standing graphene. (b) and (c) Calculated correlations between bonding energy of graphene and the corresponding core-level shift of the C\,$1s$ line, respectively, in the intercalation systems from (a) and the corresponding doping level. Data are taken from \cite{Andersen:2014fd} with permission. B: ARPES maps measured along the direction perpendicular to $\Gamma-K$ of the Brillouine zone of graphene for (a) graphene/Ir(111), (b) graphene/$0.5$\,ML\,Cs/Ir(111), and graphene/$1$\,ML\,Cs/Ir(111). Data are taken from \cite{Petrovic:2013vz} with permission. C: ARPES maps (a-c) and the corresponding constant energy cuts measured for graphene/Ir(111) before and after intercalation with oxygen. Data are taken from \cite{Larciprete:2012aaa} with permission.}
\label{gr-met_INT}
\end{figure*}

The recent low-temperature ARPES experiments ($40$\,K) on graphene/Ni(111) and graphene/Co(0001) were focused on the investigation of the energy region where \textit{hybridization} of Ni\,$3d$ and graphene\,$\pi$ states occurs~\cite{Varykhalov:2012fj}. These results suggest the existence of \textit{intact Dirac cones} in these graphene-metal systems via over-doping of graphene which is contact with Ni (Co). This consideration contradicts the present description of these systems and these observations require further experimental and theoretical analysis.

The electronic structure of graphene/Ir(111) is shown in Fig.~\ref{ARPES_grNi_grIr}(b-c), where one can clearly identify graphene-derived $\pi$ and $\sigma$ bands~\cite{Pletikosic:2009,Rusponi:2010,Kralj:2011kq,Starodub:2011a,Varykhalov:2012ec,Larciprete:2012aaa}. In this system graphene is $p$-doped and Dirac cone is found $\approx100$\,meV above $E_F$. Due to the weak interaction between graphene and Ir(111), the energy bands of graphene are not disturbed via intermixing with Ir bands. This effect is clearly indicated by the fact that the Rashba-split Ir $p_z$ surface resonance (SR) around the $\Gamma$-point remains intact upon covering Ir(111) surface with graphene (only upward shift was detected)~\cite{Kralj:2011kq,Varykhalov:2012ec} (confirmed by the XPS studies of the surface core-level shifts between Ir(111) and gr/Ir(111)~\cite{Lacovig:2009}). As was shown above, graphene on Ir(111) forms a moir\'e structure with periodicity $(10\times10)$gr/$(9\times9)$Ir(111) [Figs.~\ref{gr-met-structures}(a), \ref{gr-met_STM}(d), and \ref{STM_grRu_grIr}] and effect of this additional periodicity with small modulating potential is visible in ARPES maps as additional photoemission replicas for the main $\pi$ and $\sigma$ emission bands [Fig.~\ref{ARPES_grNi_grIr}(b-c)]. Intersections of these replica bands and main bands produces the so-called mini-gaps in the electronic structure of graphene due to the avoid-crossing mechanism. For example such replica bands and the corresponding mini-gaps at the binding energies of $\approx1$\,eV and $\approx2.5$\,eV are clearly visible in Fig.~\ref{ARPES_grNi_grIr}(c). The similar behaviour of the valence band states of graphene was found also for other graphene-metal systems with weak overlap of the valence band states at the interface: graphene/Cu(111)~\cite{Walter:2011fj,Jeon:2013ek,Avila:2013fg}, graphene/Pt(111)~\cite{Sutter:2009a,Pisarra:2014gn}, graphene/Au(111)~\cite{Wofford:2012fp}.

\section{Graphene hetero- and nano-structures on metals}

The electronic structure of graphene on metals can be tailored in different ways that might help to understand the effects observed in the real electron- and spin-transport graphene-based devices. There are many ways for such modifications and several of them will be considered here: (i) intercalation of different species between a graphene layer and metallic substrate, (ii) adsorption of atoms, molecules and clusters on top of the graphene-metal system, and (iii) preparation of graphene objects of reduced dimensionality (flakes, quantum dots, nanoribbons).

\subsection{Intercalation}

There are many recent experimental and theoretical works where different species were intercalated between graphene and metallic substrate with the aim to perform controllable modifications of the crystallographic and electronic structures of the system. Among these materials are: (i) alkali and alkali-earth metals, which induce $n$-doping of graphene~\cite{Nagashima:1994,Gruneis:2008,Bianchi:2010cu,Petrovic:2013vz,Schumacher:2013hl,Fedorov:2014aa}, (ii) \mbox{$sp$-,} $d$-, and $f$-metals, where doping depends on the intercalated metal~\cite{Shikin:1998,Farias:1999,Shikin:1999a,Farias:2000,Shikin:2000b,Shikin:2000,Dedkov:2001,Dedkov:2003,Dedkov:2008e,Varykhalov:2008,Enderlein:2010,Kang:2010,Varykhalov:2010a,Sun:2010,Weser:2011,Voloshina:2011NJP,Sicot:2012,Generalov:2012,Meng:2012ee,Lizzit:2012hh,Generalov:2012gi,Gyamfi:2012eg,Adamska:2012ef,Rougemaille:2012bu,Decker:2013ch,Pacile:2013jc,Papagno:2013ew,Rybkina:2013cn,Jin:2013es,Shikin:2013fr,Schumacher:2013ge,Zhang:2013bw,Schumacher:2013hl,Li:2013hl,Leicht:2014jy,Vita:2014aa,Voloshina:2014iy,Schumacher:2014df,Vlaic:2014gx}, and (iii) molecular species, which decouples graphene from the metallic support, like oxygen~\cite{Sutter:2010a,Larciprete:2012aaa,Granas:2012cf,Jang:2013cn,Sutter:2013kw,Jolie:2014ev}, CO~\cite{Granas:2013tl}, and C$_{60}$~\cite{Rutkov:1995,Shikin:2000a,Varykhalov:2010}.

The intercalation of monovalent species (alkali atoms and halogens) as well as oxygen and CO in graphene/Ir(111) was studied theoretically in Ref.~\cite{Andersen:2014fd} [Fig.~\ref{gr-met_INT}(A)]. It was shown that these materials effectively decouple a graphene layer from the metallic substrate and induce the corresponding doping of graphene. For example alkali metals induce $n$-doping and oxygen or halogens make graphene $p$-doped. Authors of Ref.~\cite{Andersen:2014fd} found a correlation (linear dependence) between bonding of graphene to substrate and the doping level of graphene [Fig.~\ref{gr-met_INT}(A,a-b)]. This fact indicates that for the studied species (simple case of the charge transfer) the interaction between graphene and intercalants has ionic character. Authors also found a linear dependence of the shift of the C\,$1s$ line on the doping level.

The results of this theoretical work are confirmed by the experimental ARPES data shown in Fig.~\ref{gr-met_INT} for (B) graphene/Cs/Ir(111)~\cite{Petrovic:2013vz} and (C) graphene/O/Ir(111)~\cite{Larciprete:2012aaa}. One can clearly see that in both cases graphene is effectively decoupled from Ir(111) as the corresponding replica bands induced by the moir\'e structure of gr/Ir(111) are not visible in the respective intercalation systems. The obtained doping level of $E-E_F=-1$\,eV for gr/Cs/Ir(111) [Fig.~\ref{gr-met_INT}(B)]~\cite{Petrovic:2013vz} and $E-E_F=+0.64$\,eV [Fig.~\ref{gr-met_INT}(C)]~\cite{Larciprete:2012aaa} correspond to the strongly $n$- and $p$-doped graphene, respectively, and these values are in very good agreement with the doping levels obtained in DFT calculations [Fig.~\ref{gr-met_INT}(A,a)]~\cite{Andersen:2014fd}.

\begin{figure*}[t]
\includegraphics[width=\linewidth]{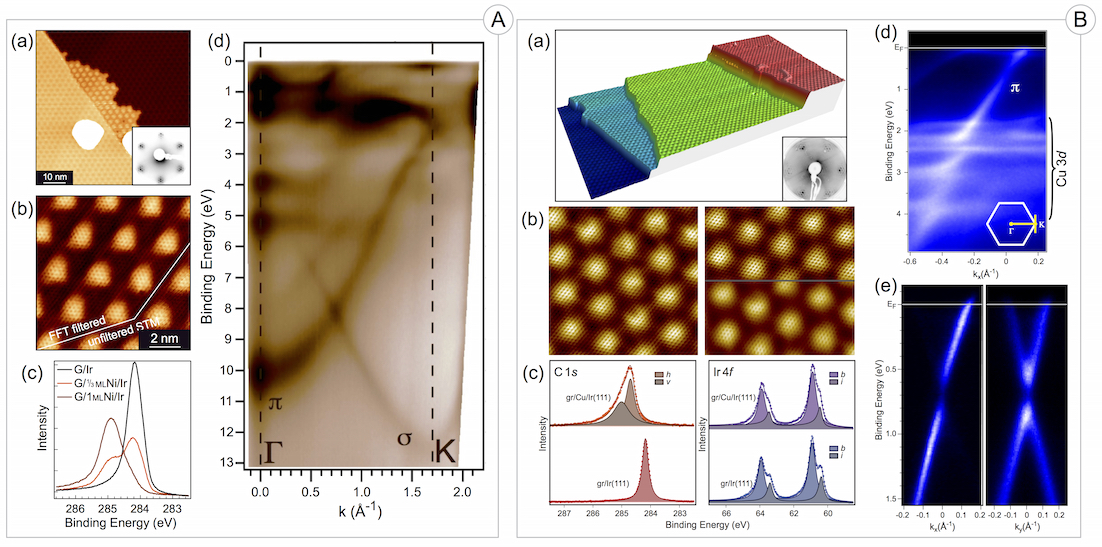}
\caption{A: (a-b) Large and small scale STM images of graphene/Ni/Ir(111) demonstrating the process of intercalation and the structure of the system. Inset of (a) shows the corresponding LEED. (c) Comparison of C\,$1s$ XPS spectra at different stages of intercalation of Ni in graphene/Ir(111). (d) ARPES map for graphene/Ni/Ir(111) measured along the $\Gamma-K$ direction. Data are taken from \cite{Pacile:2013jc} with permission. B: Large and small scale STM images of graphene/Cu/Ir(111). Inset of (a) shows the corresponding LEED. (c) Comparison and the respective fit of C\,$1s$ and Ir\,$4f$ XPS spectra of graphene/Ir(111) and graphene/Cu/Ir(111). (d) ARPES map on the large energy scale and (e) the corresponding zoom to the energy range close to $E_F$ for graphene/Cu/Ir(111). Data are taken from \cite{Vita:2014aa} with permission.}
\label{grNiIr_grCuIr}
\end{figure*}

Intercalation of graphene on the particular metallic substrate might help to synthesise the artificial system which cannot be prepared in a direct way. Such examples are shown in Fig.~\ref{grNiIr_grCuIr}(A,B). In the first case graphene/Ir(111) was intercalated with Ni and strongly corrugated graphene, compared to weakly buckled graphene/Ir(111), was obtained~\cite{Pacile:2013jc}. Nickel layer was deposited at room temperature on top of graphene/Ir(111) and then system was annealed at temperatures up to $800$\,K. It was found that Ni intercalates under graphene as shown by STM and core-level XPS [Fig.~\ref{grNiIr_grCuIr}(A,a-c)]. In this case Ni grows in a pseudomorphic way at the interface between graphene and Ir(111) that periodicity and symmetry of the system is preserved [LEED and STM in Fig.~\ref{grNiIr_grCuIr}(A,a-c)]. As was found in STM the imaging contrast in this case is changed and STM images are very similar to those for the strongly buckled graphene/Ru(0001) system. In the same experiments with identical imaging conditions (bias voltage and tunneling current), the corrugation of $0.6$\,\AA\ was measured for gr/Ni/Ir(111) compared to $0.25$\,\AA\ for gr/Ir(111). DFT simulation of gr/Ni/Ir(111) yields the corrugation of graphene of $1.51$\,\AA\ [$1.195$\,\AA\ for gr/Ru(0001)] with a minimal distance between graphene and Ni-layer of $1.94$\,\AA\ ($2.195$\,\AA\ for gr/Ru(0001) and $2.1$\,\AA\ for gr/Ni(111)). It is interesting that more than $70$\% of carbon atoms are placed at the distance $2.0-2.2$\,\AA\ above Ni layer giving as a results the STM images similar to those for gr/Ru(0001).

XPS spectra measured before and after Ni-intercalation in gr/Ir(111) demonstrate the strong modification of the C\,$1s$ line [Fig.~\ref{grNiIr_grCuIr}(A,c)]: the single peak at $284.1$\,eV for gr/Ir(111) is shifted to $284.9$\,eV for gr/Ni/Ir(111) and shows a strong asymmetry after intercalation indicating the existence of two regions in a graphene layer where weak and strong orbital mixing are observed [similar to gr/Ru(0001) where double peak structure of C\,$1s$ was found, Fig.~\ref{NEXAFS_gr-met}(B,f)].

The band structure of the graphene/Ni/Ir(111) system measured by means of ARPES is shown in Fig.~\ref{grNiIr_grCuIr}(A,d) and it is very similar to the one for graphene/Ni(111): the graphene-derived $\pi$ states are found at $\approx10$\,eV of the binding energy at the $\Gamma$-point and the Ni\,$3d$ weakly dispersing bands are located just below $E_F$. The position of the $\pi$ band at the $K$-point depends on the thickness of the intercalated Ni layer: it is changed from $2.16$\,eV for 1\,ML-intercalated-Ni to $2.65$\,eV for multilayer-intercalated Ni~\cite{Pacile:2013jc} (this effect is connected with the narrowing of the $d$-band with decrease of the thickness of the Ni layer and demonstrate the effect of Ni\,$3d$ -- graphene\,$\pi$ orbital overlap or \textit{hybridization}). These results support the model discussed above (see Sec.~\ref{ELSTR} and Ref.~\cite{Voloshina:2014jl}) where modification of the electronic structure of graphene on the open $d$-shell metal depends on the relative energy position of the shifted Dirac cone and the $d$-band: complete space-, energy-, and wave-vector-overlap leads to the destroying of the electronic structure of graphene around $E_F$ with formation of the so-called \textit{hybrid} states below $E_F$.

Another example of the artificial system, graphene/ Cu/Ir(111), obtained after intercalation, is shown in Fig.~\ref{grNiIr_grCuIr}(B)~\cite{Vita:2014aa}. Similar to the previous system, the copper layer grows pseudomorphically underneath graphene as deduced from LEED and STM data [Fig.~\ref{grNiIr_grCuIr}(B,a-b)]. Contrary to graphene/Ir(111), the obtained intercalation system gr/Cu/Ir(111) was always imaged in STM in the \textit{true} topographic contrast and its variation upon changes of the bias voltage was not detected. This system was modelled by DFT (PBE-D2) and these calculations gave very good agreement with the experimental STM images. Relaxation of the geometry of graphene/Cu/Ir(111) yields a corrugation of graphene of $0.229$\,\AA\ with a mean distance between graphene and Cu layers of $3.02$\,\AA. These calculations also predict the $n$-doping of graphene with the position of the Dirac point at $0.45$\,eV below $E_F$. 

XPS data for gr/Cu/Ir(111) confirm these predictions of the DFT calculations and consistent with STM data [Fig.~\ref{grNiIr_grCuIr}(B,c)]. The C\,$1s$ XPS line is shifted to the larger binding energies and can be fitted with two components which ratio gives the ratio of the areas for \textit{hills} and \textit{valleys} of the gr/Cu/Ir(111) structure. The respective changes are also observed for the Ir\,$4f$ XPS line: the emission from the interface component ($i$) is suppressed for gr/Cu/Ir(111) compared to gr/Ir(111) and its binding energy is increased indicating that chemical surrounding for these Ir atoms is also changed.

ARPES intensity maps for graphene/Cu/Ir(111) are shown in Fig.~\ref{grNiIr_grCuIr}(B,d-e). They show an $n$-doping of graphene in this system of about $0.688$\,eV. Also the clear \textit{hybridization} between Cu\,$3d$ and graphene\,$\pi$ bands is visible in the energy range of $\approx2-4$\,eV that leads to the opening of the energy gaps due to the avoided-crossing mechanism. Surprisingly, a large energy gap of $0.36$\,eV is opened directly at the position of the Dirac point. DFT calculations (PBE-D2) give a value of $0.15$\,eV. Similar behaviour of graphene valence band states and gap opening at the Dirac point was also observed in Refs.~\cite{Varykhalov:2010a,Papagno:2013ew} for other graphene-metal intercalation systems (Cu, Ag, Au as intercalants). The proposed in Ref.~\cite{Varykhalov:2010a} explanation that value of the gap depends on the doping level of graphene does not supported by other ARPES data (for example, the absence of the energy gap for alkali-metal intercalated graphene-metal interface~\cite{Petrovic:2013vz,Fedorov:2014aa}). Authors of Refs.~\cite{Vita:2014aa,Voloshina:2014jl} propose a model that appearing of the energy gap at the Dirac point in the electronic structure of graphene, adsorbed on the open $d$-shell metal, is due to the violation of the symmetry of the electronic states as the $p_z$ orbitals of carbon atoms from different sublattices of graphene overlap with the $d$ orbitals of the different symmetries of the same metal atom below a graphene layer. This model was tested for different doping levels of graphene and different distances between graphene and surface of $d$-metal and clear correlation between these parameters and the width of the energy gap was found~\cite{Voloshina:2014jl}. These changes have a huge impact on the electronic structure of graphene around the Dirac point and might influence the transport properties of graphene at the graphene-metal interfaces.

\subsection{Adsorption on graphene/metal}

\begin{figure*}[t]
\includegraphics[width=\linewidth]{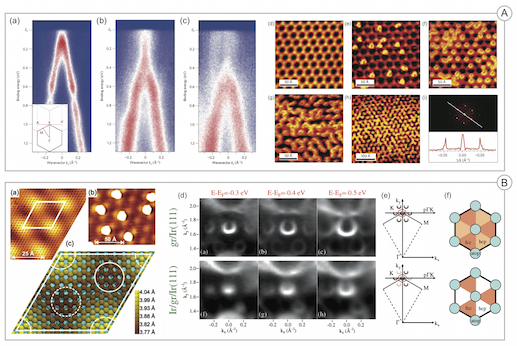}
\caption{A: (a-c) ARPES maps recorded along the $A-K-A$ direction (see inset of (a)) for $0$\,sec, $30$\,sec, and $50$\,sec, respectively, of dosing of atomic hydrogen on graphene/Ir(111). (d-h) STM images of graphene/Ir(111) taken after very small time, $15$\,sec, $30$\,sec, and $50$\,sec, respectively, of dosing of atomic hydrogen. (i) FFT analysis of image (h). Data are taken from \cite{Balog:2010} with permission. B: STM images of (a) graphene/Ir(111) and (b) $0.02$\,ML\,Ir/graphene/Ir(111). (c) Structural model of graphene/Ir(111) with C-Ir distances obtained from GGA calculations. (d) Constant energy ARPES maps for graphene/Ir(111) and $0.15$\,ML\,Ir/graphene/Ir(111). (e) Scheme of the Dirac cones for the main photoemission band and its replicas. (f) Colour coded scheme of the distribution of the moir\'e potential in the systems. Data are taken from \cite{Ndiaye:2006,Rusponi:2010} with permissions.}
\label{gr-met_ADS}
\end{figure*}

The aim of the studies of adsorption on top of graphene is manyfold. Firstly, one can expect the selective adsorption of atoms of different nature with the result that the electronic structure of graphene will be modified~\cite{Balog:2010,Ng:2010,Haberer:2010,Haberer:2011eu,Usachov:2011dl,Bottcher:2011fm,Larciprete:2012fm,Rajasekaran:2012ji}: doping, overlap of orbitals of adsorbate and valence states of graphene, and in ideal case the violation of the sublattice symmetry in graphene that can lead to the opening of the energy gap at the Dirac point. Secondly, if e.\,g. moir\'e graphene structures on metals are studied, then deposition of molecules or metals can lead to the formation of ordered arrays of molecules or clusters on top~\cite{Ndiaye:2006,Feibelman:2008,Pan:2009,NDiaye:2009a,Donner:2009,Feibelman:2009,Rusponi:2010,Sicot:2010,Liu:2011en,VoVan:2011ia,Wang:2011hh,Zhang:2011ky,Engstfeld:2012fh,Knudsen:2012ei,Cavallin:2012hp,Papagno:2012hl,Bionski:2012gn,Yang:2012kq,Hamalainen:2012bx,Li:2012in,Zhang:2012il,Bazarnik:2013gl,Lei:2013jq,Garnica:2013ig,Han:2013ec,Franz:2013gs,Jarvinen:2014va}. In this case the properties of the complete system can be modelled on the basis of its single element -- single cluster -- and different properties of such objects (catalytic activity or magnetic properties for magnetic elements) can be studied in details.

Figure~\ref{gr-met_ADS}(A) shows an example of adsorption of atomic-H on graphene/Ir(111)~\cite{Balog:2010,Ng:2010}. The panels (A,a-c) show the evolution of the electronic structure of graphene around the $K$-point of the Brillouine zone as a function of the exposure time of the system to atomic hydrogen (these maps show the electronic structure of graphene measured along the direction perpendicular to $\Gamma-K$, see also Fig.~\ref{ARPES_grNi_grIr}(b,c)). The clear modification of the emission picture is visible which indicated by opening of the energy gap at $E_F$ for higher exposure time, broadening of the emission lines, and reduction of the photoemission intensity. The parallel STM measurements performed in the similar experimental conditions [Fig.~\ref{gr-met_ADS}(A,d-i)] demonstrate the selective hydrogen adsorption at the FCC high-symmetry places of the graphene/Ir(111) moir\'e structure on the initial stages of the experiment; increase of the adsorption time leads to the formation of the elongated structures as observed in STM. However, even long hydrogen dosing allows to identify that adsorption of H-atoms on graphene/Ir(111) follows its moir\'e structure [see FFT analysis in Fig.~\ref{gr-met_ADS}(A,i)].

DFT analysis of the hydrogen adsorption on the graphene/Ir(111) structure performed in the same work \cite{Balog:2010} shows that it is energetically favourable for hydrogen to form a ``graphane-like'' islands at the FCC places of gr/Ir(111), where one of the C-atoms from the unit cell of graphene (C-$hcp$) is bonded to the H atom and the second C-atom (C-$top$) is bonded to the underlying Ir atom. Such adsorption configuration leads to the local rehybridization of the carbon atoms in the graphene layer from $sp^2$ to $sp^3$. Such regions of the ``graphane-like'' structures have very large band gaps (might reach $3.5$\,eV for \textit{true} graphane~\cite{Sofo:2007}). In reality, DFT calculations show that, e.\,g. at $23$\% coverage of graphene with atomic hydrogen, a band gap of $0.73$\,eV can be obtained. As was concluded on the basis of the experimental and theoretical data, the band gap opening in the H-graphene/Ir(111) system is caused by the confinement effect in the residual bare graphene regions and the broadening of the bands is due to the increased uncertainty in the wave-vector for electrons in such structure~\cite{Balog:2010}.

The investigations of the cluster formations on top of graphene/metal moir\'e structures and their influence on the electronic structure of graphene were performed in a series of experimental and theoretical works listed above. The first experiments were carried out for Ir-clusters on graphene/Ir(111) [Fig.~\ref{gr-met_ADS}(B,a-c)]~\cite{Ndiaye:2006}. These STM results combined with DFT calculations show that Ir atoms preferentially nucleate at the HCP high-symmetry regions of graphene/Ir(111) forming a regular arrays of Ir clusters. According to DFT calculations adsorption of Ir atoms on graphene/Ir(111) leads to the local rehybridization (from $sp^2$ to $sp^3$) of carbon atoms at the FCC and HCP regions (similar to discussed earlier). In this case the C-atom which is placed at the $hcp$ or $fcc$ position is covalently bound to the Ir atom placed above the graphene layer and the C-atop at the $top$ position is covalently bound to the underlying Ir atom~\cite{Ndiaye:2006,Feibelman:2008,Feibelman:2009}. The similar situation is observed for other metallic adsorbates~\cite{NDiaye:2009a,Knudsen:2012ei,Franz:2013gs}.

Adsorption of clusters or molecules on the graphene/metal systems can lead to the changes in the electronic structure of a graphene layer (change of the doping level or/and modification of the energy dispersion of the graphene valence band states) as was demonstrated in the experiment. The electronic structure of the Ir/graphene/Ir(111) system was studied by means of ARPES and these results are shown in Fig.~\ref{gr-met_ADS}(B,d-f)~\cite{Rusponi:2010}. As was found in the experiment the deposition of Ir on gr/Ir(111) (the nominal thickness of Ir was $0.15$\,ML that corresponds to one $13$-atom Ir cluster per HCP region of gr/Ir(111)) leads to the opening of the energy gap in the electronic structure of graphene at the $K$-point and position of the lower part of the Dirac cone is placed by $200\pm20$\,meV below $E_F$. The important result is that the symmetry of the system is reduced from six-fold to three-fold as can be deduced from the comparison of photoemission maps for gr/Ir(111) and Ir/gr/Ir(111) [Fig.~\ref{gr-met_ADS}(B,d-e)]. This reflects the fact that local rehybridization from $sp^2$ to $sp^3$ appears at the HCP regions upon Ir clusters formation and this perturbs the photoemission intensity suppressing three out of the six photoemission replicas~\cite{Rusponi:2010}. The formation of the Ir arrays of clusters on gr/Ir(111) also increases the width of the mini-gaps where replica bands cross the main graphene $\pi$-band from $240\pm20$\,meV for gr/Ir(111) to $330\pm20$\,meV for Ir/gr/Ir(111) indicating the effect of the strengthening of the modulating moir\'e potential. As a result the strong anisotropy of the group velocity of the graphene $\pi$ states along $\Gamma-K$ and perpendicular to this direction was measured that indicates the importance of the periodic potential on the transport properties of graphene and can help to tailor the transport properties of graphene in the future devices. The discussed effects (doping and increasing of the width of mini-gaps) were dramatically increased by the co-adsorption of Ir and Na atoms on graphene/Ir(111)~\cite{Papagno:2012hl}, that was explained by the even stronger modulations of the moir\'e potential in the obtained systems.

Magnetic properties of magnetic clusters on top of graphene/Ir(111) were studied by means XMCD in Ref.~\cite{VoVan:2011ia}. Different clusters were studied: Pt$_{13}$Co$_{26}$, Pt$_{13}$Fe$_{26}$, Ir$_{13}$Co$_{26}$, Ir$_{50}$Co$_{500}$, Co$_{2700}$. The morphology and quality of the studied systems were verified by STM. It was found that for small clusters, e.\,g. Pt$_{13}$Co$_{26}$, magnetization curves do not reach the saturation even at magnetic fields of $5$\,T and no hysteresis down to $10$\,K was observed. The extracted spin and orbital magnetic moments of Co in such clusters are $\mu_s=1.5\pm0.2\mu_B$ and $\mu_l=0.22\pm0.02\mu_B$, respectively ($\mu_s=1.62\mu_B$ and $\mu_l=0.15\mu_B$ for bulk Co). Larger clusters (Ir$_{50}$Co$_{500}$ and Co$_{2700}$) demonstrate the non-zero coercivity, which vanishes around $40$\,K. As was found, the spin (orbital) magnetic moments for such clusters are slightly increased (decreased)  by $0.2\mu_B$ ($0.02-0.04\mu_B$) compared to the values for small clusters.

In Ref.~\cite{Gerber:2013fa} the regular array of Pt clusters was formed on graphene/Ir(111). This system was exposed to CO gas and its stability was tested by means of STM, XPS, and results were analysed with DFT calculations. It was found that for the clusters of the size of few tens of atoms such adsorption of CO leads to the sintering of the clusters via Smoluchowski ripening -- cluster diffusion and coalescence. Larger clusters upon exposure to CO remain stable but form three-dimensional larger agglomerates. These effects were explained by the weakening of the Pt-C interaction upon CO adsorption in the 2-fold edge bridge positions of the cluster between Pt atoms. As was found such position is energetically more favourable compared to the 1-fold position when CO is bonded to every Pt edge atom of the cluster. Such 2-fold adsorption of CO leads to the increase of the Pt-C distance by $0.9$\,\AA\ and clusters become less bonded to the gr/Ir(111) substrate that increase their mobility and probability to coalescence to large structures. Such studies of the stability of the cluster arrays on graphene/metal systems might shed light on the understanding of their catalytic properties in future studies.

\begin{figure}[t]
\includegraphics[width=\linewidth]{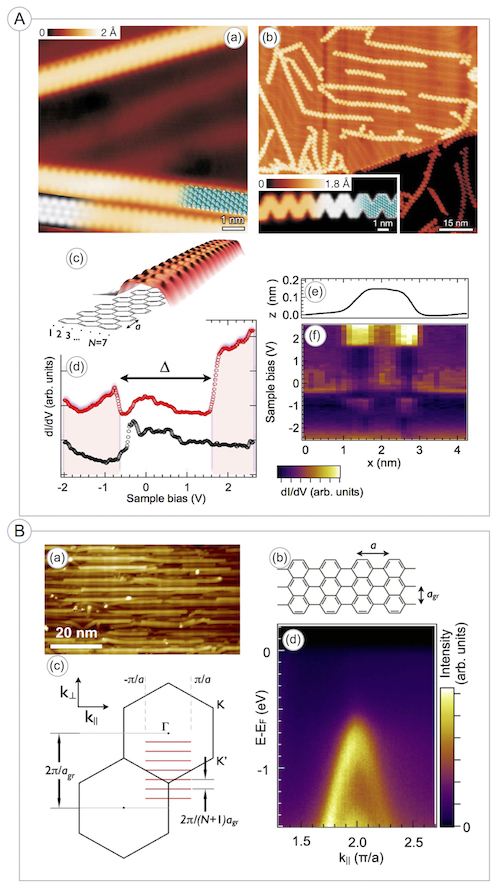}
\caption{A: STM images of (a) 7-AGNRs and (b) chevron-type GNRs on Au(111). In both panels the STM images are overlaid with DFT-based simulations of STM (grey) and atomic models of GNRs (C: blue, H: white). (c) 3D representation of STM image of 7-AGNR. (d) STS spectra of 7-AGNR (red) and clean Au(111) (black). (e) STM profile across 7-AGNR. (f) Series of colour-coded STS spectra taken across 7-AGNR. Data are taken from \cite{Cai:2010,Ruffieux:2012et} with permissions. B: (a) STM image of unidirectionally aligned 7-AGNRs on Au(788). (b) Structure of hydrogen-terminated 7-AGNR and relevant lattice parameters. (c) Brillouin zone of graphene (black hexagon) and the one- dimensional Brillouin zone of 7-AGNR (red). (d) ARPES intensity plot $I(E-E_F,k_{||})$ recorded along the ribbon axis. Data are taken from \cite{Ruffieux:2012et} with permissions.}
\label{gr-met_NANO_1}
\end{figure}

\subsection{Graphene nanoobjects: nanoribbons, nanoflakes, quantum dots}

A graphene nanoribbon (GNR) is a narrow strip of graphene, which structure and the electronic properties are defined by the edge morphology and the width. The morphology is defined by the chiral indexes ($n,m$) (similar to CNTs) or chiral angle $\theta$. GNRs can be classified, with respect to the morphology of edges, as armchair, zigzag or chiral. The electronic structure of GNRs is different from the one of graphene due to the confinement effects and as was shown theoretically, depending on the edge morphology and the width, GNRs can be metallic, semimetallic or semiconducting~\cite{Nakada:1996us}. According to these considerations~\cite{Nakada:1996us,Fujita:1996vs,Wakabayashi:1999ti}, zigzag GNRs (ZGNRs) show a sharp peak in the density of the electronic states at $E_F$ that leads to a net spin polarization at the edges; antiferromagnetic coupling between two edges opens a small fundamental gap. The armchair graphene nanoribbons (AGNRs) can be either semimetallic or semiconducting depending on the width of the nanoribbon~\cite{Brey:2006cb,Sasaki:2006wr,Abanin:2006ds,Lee:2005ek,Son:2006ky}.  

However, the contact of graphene nanoribbon with a metallic surface can drastically change the electronic structure of GNR due to the effects of doping and orbital overlap of the electronic states of nanoribbon and metal as was shown above for a graphene-metal interfaces. Thus, the previous theoretical consideration have to be revisited with the aim to account for the metallic contact to GNRs. For example in Ref.~\cite{Archambault:2013fd} it was shown that although the interaction of nanoribbons with noble-metals is weak, adsorption of GNRs on Pd or Ti leads to the strong orbital mixing of the electronic states at the place of contact. The DFT calculations performed in this work show that the so-called metal-induced gap states appear in the energy gap of GNR at the interface between nanribbon and metal. These states can effectively penetrate on the large distance inside GNRs that can lead to the shortening of the metallic contacts and this effect limits the application of small GNRs in the devices. 

Presently GNRs can be prepared even with atomic precision in different ways~\cite{Han:2007bl,Li:2008ht,Wang:2010cf,Datta:2008if,Ci:2008je,Campos:2009cd,Kosynkin:2009bw,Jiao:2010de,Tao:2011kl,vanderLit:2013jf}. Fig.~\ref{gr-met_NANO_1}(A) shows the STM images of (a) AGNRs and (b) chevron-type GNRs, respectively, synthesised on Au(111) via \textit{bottom-up} approach from different molecular precursors~\cite{Cai:2010}. STS measurements performed on these nanoribbons give a bandgap of $2.3\pm0.1$\,eV for 7-AGNR [see Fig.~\ref{gr-met_NANO_1}(A,c-d)]~\cite{Ruffieux:2012et} and $1.4\pm0.1$\,eV for 13-AGNRs~\cite{Chen:2013fa}. These values can be compared with the gap of $2.3-2.7$\,eV as deduced from the quasiparticle $GW$ calculations corrected for the image charge in metals~\cite{Ruffieux:2012et}.

Later the combined ARPES and inverse photoemission (IPES) measurements were performed on these GNRs [ARPES data for 7-AGNRs: Fig.~\ref{gr-met_NANO_1}(B); IPES data: not shown]~\cite{Ruffieux:2012et,Linden:2012fu}. In this case the alignment of GNRs on the macroscopic scale is required and the stepped Au(788) surfaces were used for this purpose. This surface consists of the $3.83$\,nm-wide $\{111\}$ terraces which can be used for the template growth of both types GNRs, straight ANRs and chevron-type. These experiments yield $2.8\pm0.4$\,eV, $1.6\pm0.4$\,eV, and $3.1\pm0.4$\,eV for 7-AGNRs, 13-AGNRs, and chevron-type GNRs, respectively. Modelling of these GNRs on Au(111) within DFT and many-body electron approaches gives a band gap of $2.85$\,eV and $2.96$\,eV for 7-AGNRs and chevron-type GNRs, respectively~\cite{Liang:2012wi}, which are in rather good agreement with experimental values. These calculations indicate the charge transfer from GNRs on Au(111) that leads to the surface polarization, that influences the width of the energy gap of GNRs.

Later the above presented method of synthesis of GNRs~\cite{Cai:2010} was used in Ref.~\cite{vanderLit:2013jf} where structural and electronic properties of nanoribbons and their dependence on the edge termination were studied via combination of STM and AFM with the CO-terminated scanning tips. It was found that initially synthesised GNRs have H-terminated edges and in this case the bonding to the Au(111) substrate is weak that allows to move them with the STM tip. At the same time the STS measurements at the edge of GNR show the existence of the so-called vibronic tunnelling mode. In further experiments, the structure of GNR was modified via removing one of the H-atoms at one of the its end as was clearly identified via combined STM/AFM measurements -- the \textit{bond formation} between edge C-atom and the underlying Au was found. This modification leads to the blocking of the mobility of GNR on the surface and also to the drastic changes in the STS spectra. In this case the STS spectrum measured at the modified edge shows no fine structure (only very broad peak) and the one measured at the non-modified edge demonstrate the strong suppression of the vibronic mode. These results demonstrate the importance of the edge termination and the possible formation of the GNR-metal contacts for the modelling and interpretation of the transport properties of nanoribbons. 

\begin{figure*}[t]
\includegraphics[width=0.9\linewidth]{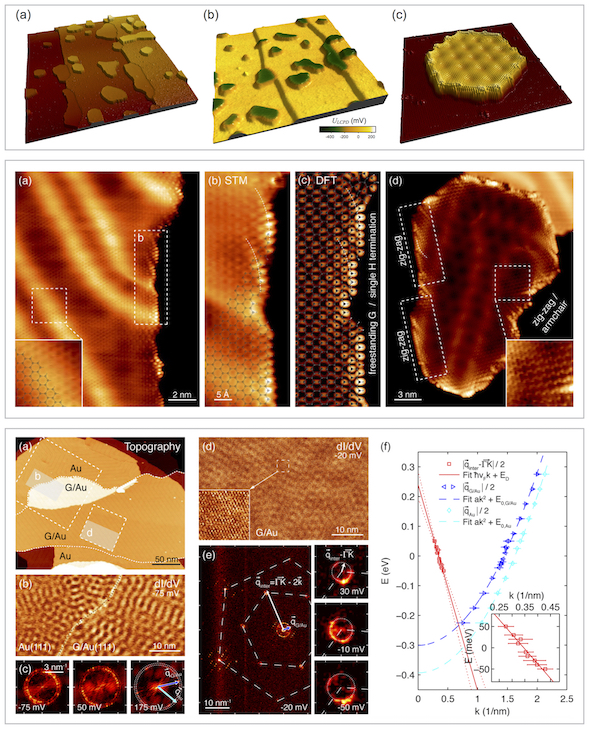}
\caption{Upper panel: 3D representation of data for GNFs and GNDs on Ir(111) obtained in (a) STM ($180\times180\mathrm{nm}^2$) and (b) combined AFM/KPFM ($150\times150\mathrm{nm}^2$) measurements. In (b) the topography of the system measured by AFM is overlaid by the respective KPFM signal measured simultaneously. (c) Atomically resolved STM image ($20.2\times20.2\mathrm{nm}^2$) of the single GQD on Ir(111). Middle panel: (a,b,d) experimental and (c) theoretical STM images of the H-edge-terminated graphene nanoflakes on Au(111). Bottom panel: (a) STM topography of an approximately $400\times160$\,nm$^2$ large flake used for $dI/dV$ mappings. (b) $dI/dV$ map of a large graphene flake marked by ``b'' in (a). (c) FFT images at selected bias voltages. (d) $dI/dV$ map on the graphene flake with atomic resolution. (e) FFT maps of the image (d) at different bias voltages. (f) Plots of the Au(111)- and graphene-related dispersions of the electronic states. The $k$ values are plotted with respect to the $\Gamma$-point in the case of the surface state and with respect to the $K$-point in case of graphene. Data are reproduced from Ref.~\cite{Leicht:2014jy} with permission.}
\label{gr-met_NANO_2}
\end{figure*}

There are two common ways to synthesise graphene nanoflakes (GNFs) or quantum dots (GQDs) on metallic surfaces: (i) the so-called temperature programmed growth (TPG) when on the first step the hydrocarbons (C$_2$H$_4$~\cite{Coraux:2009,Hamalainen:2011ja,Subramaniam:2012fp} or coronene~\cite{Coraux:2009}) are predeposited on the metallic surface and then this system is annealed at elevated temperature or (ii) unfolding the predeposited C$_{60}$ molecules as demonstrated in Ref.~\cite{Lu:2011bg}. The upper panel of Fig.~\ref{gr-met_NANO_2} shows 3D images of graphene flakes and quantum dots obtained in STM (a,c) and in combined AFM/KPFM (b) measurements. These GQDs were formed on Ir(111) via TPG method and C$_2$H$_4$ was used as a precursor. Combined AFM/KPFM measurements [Fig.~\ref{gr-met_NANO_2}, upper panel, (b)] yield the LCPD value of $\approx600$\,meV, which can be compared with the measured work function difference between graphene and Ir(111) of $1.6$\,eV and $1.1\pm0.3$\,eV obtained from LEEM~\cite{Starodub:2012cb} and STS~\cite{Forster:2012aa} experiments, respectively. The observed discrepancy can be assigned to the smearing effect of the macroscopic scanning tip during KPFM measurements. This synthesis method of graphene nano-dots (GNDs) leads to the formation of well formed graphene islands with straight edges oriented along main crystallographic directions of the Ir(111) surface. The high quality of such islands allows the simultaneous atomically-resolved STM/AFM imaging of graphene and metallic surface [Fig.~\ref{gr-met_NANO_2}, upper panel, (c)] giving a possibility to carefully trace the crystallographic and electronic structure of nanostructures on the atomic scale.

The electronic structure of such GNDs on Ir(111) was extensively studied by STS in several recent works~\cite{Hamalainen:2011ja,Phark:2011de,Subramaniam:2012fp,Altenburg:2012kp}. These GQDs have exclusive zigzag edges. The presence of edge-states was not detected~\cite{Li:2013ji} as supported by DFT calculations and it was explained by a hybridisation of the GQDs $p_z$ orbitals and the substrate valence band states (here: Ir $5d_{z^2}$ surface state). Such interaction gradually decreases in strength from the edge towards the centre of the GQD. It is interesting to note, that although these works present the similar experimental observations, they do not provide a clear explanation of these effects (effects of quantization as well as the explanation for the extracted dispersion of the electronic states, $E(k)$) and did not give an answer about contributions of the electronic states of graphene and metallic substrate in the tunnelling current and, hence, in the observed effects.

The properties of such GNDs can be tailored in different ways (intercalation, edge termination, manipulations, etc.). For example in Refs.~\cite{Leicht:2014jy,Jolie:2014ev} the GQDs/Ir(111) was intercalated either with Au or oxygen, respectively. In the first case, Ref.~\cite{Leicht:2014jy}, the thick layer of gold ($50-100$\,\AA) was deposed on GQDs/Ir(111) prepared by TPG and then system was annealed that leads to the formation of graphene flakes on Au(111). The quality of this flakes is very high as can be depicted from Fig.~\ref{gr-met_NANO_2} (middle and bottom panels). These STM images (middle panel, a) give a possibility to simultaneously resolve a herringbone structure of Au(111), moir\'e structure of the graphene/Au(111) interface (as due to the lattice mismatch between two materials), atomic contrast of the graphene layer, and the edge-scattering effects in the GND. Comparison of the experimental STM data and the results obtained in DFT simulations (middle panel, b-c) demonstrates very good agreement for the H-terminated GNFs. The hydrogen termination and the weak interaction between graphene and Au(111) are supported by the possibility to move flakes with the scanning tip as it was demonstrated in the experiment. Such possibility is absent for the graphene nanoflakes on Ir(111) due to the C-Ir bonding at the edges as was demonstrated in Ref.~\cite{Lacovig:2009}. 

The STS measurements performed on the GNFs/Au(111)/Ir(111) system allow to separate contributions in the tunneling current from graphene and from the substrate. Such experiments were performed on GNF, which is marked in the STM image shown in Fig.~\ref{gr-met_NANO_2} (bottom panel, a) by letter ``b'', and its STS map measured at the bias voltage of $-75$\,mV is shown in (b). These data show the characteristic standing wave patterns and the corresponding ring structure in the FFT images obtained from the measured data at the different bias voltages (bottom panel, c) is unambiguously assigned to the surface state of Au(111). The obtained dispersion of the electronic states, $E(k)$, where $q_{Au,gr/Au}=2k$, has a parabolic dispersion (bottom panel, f) with the effective mass $m^*=0.26m_e$.

Analogous STS measurements performed with atomic resolution [Fig.~\ref{gr-met_NANO_2} (bottom panel, d-e)] show in the FFT images additional structure: (i) six spots corresponding to the reciprocal lattice of graphene, which are superposed by the spots originating from the reconstruction of Au(111) and the moir\'e structure (large hexagon); (ii) ring-like features which build a hexagon corresponding to the $(\sqrt{3}\times\sqrt{3})R30^\circ$ structure in the real space and related to the intervalley scattering ($\vec{q}_{\mathrm{inter}}=\Gamma K-2\vec{k}$). These data obtained for different bias voltages allow to plot the dispersion of these electronic states [red squares in (f)]. The linear fit of these data gives the Fermi velocity of $v_F=(1.1\pm0.2)\cdot10^6$\,m/s and the position of the Dirac point at $E-E_F=0.24\pm0.05$\,eV, i.\,e. graphene is $p$-doped as found in the earlier photoemission studies of the graphene/Au(111) system~\cite{Enderlein:2010}. These measurements performed on the same graphene flake show the spectroscopic features from the surface state of Au(111) and from the intervalley scattering of the graphene Dirac fermions allow to give an answer about different contributions in the imaging of graphene on metals~\cite{Hamalainen:2011ja,Phark:2011de,Subramaniam:2012fp,Altenburg:2012kp,Leicht:2014jy,Jolie:2014ev,GarciaLekue:2014ce}.

Later experiments performed on the GQDs/oxygen/ Ir(111) system~\cite{Jolie:2014ev} give the Fermi velocity of $v_F=(0.96\pm0.07)\cdot10^6$\,m/s and the position of the Dirac point at $E-E_F=0.64\pm0.07$\,eV for graphene-related states and these values agree well with the data obtained in ARPES experiments in the same and similar works~\cite{Jolie:2014ev,Larciprete:2012aaa}.

\section{Conclusions}

The present manuscript reviews the recent progress in the studies of the structure and the electronic properties of the model graphene -- metal systems. It is discussed that the properties of graphene are defined by the interplay of morphology of the system and the overlap of the electronic valence states of graphene and metal at the interface. These effects define the modifications of the dispersion of the electronic states of graphene around the Fermi level, like the shift of the Dirac cone (doping), destruction of the Dirac cone due to \textit{hybridization} of the valence band states of metal and graphene, or/and band gap opening in the electronic spectrum of the graphene $\pi$ states due to the violation of the sublattice symmetry in graphene. Considering all these effects one can conclude that electronic spectrum of graphene, which is in contact with metal, is always strongly disturbed leading to the loss by graphene its unique properties that influence the transport properties of the devices where graphene/metal junctions are present. Here we try to summarise some ideas that might help to overcome these difficulties via application of different methods that tailor the properties of the graphene/metal interfaces or via formation of nano-objects on the basis of graphene on metals. 
 
\section*{Acknowledgements}

We would like to acknowledge our colleagues for the useful discussions, in particular, M. Fonin, K. Horn, M. Weser, S. B\"ottcher, H. Vita, O. Rader, P. Leicht, A. Varykhalov, D. Pasil\'e, M. Papagno, S. Lizzit and many others. E.N.V. appreciate the support from the German Research Foundation (DFG) through the Priority Programme (SPP) 1459 ``Graphene''. 

\section*{References}

%\bibliography{/Users/YuDedkov/Work/Articles/___REFERENCES___/references_all.bib}

\providecommand{\newblock}{}

\end{document}